\documentclass[oldversion]{aa}  
\usepackage{graphicx}
\usepackage{txfonts}
\usepackage{natbib}
\usepackage{epstopdf}
\usepackage{wasysym}
\bibpunct{(}{)}{;}{a}{}{,} 
\begin{document}
\title{Spectroastrometry of rotating gas disks for the detection of supermassive black holes in galactic nuclei.}
\subtitle{II. Application to the galaxy Centaurus A (NGC 5128).}

\author{A. Gnerucci\inst{1}, 
            A. Marconi\inst{1},
           A. Capetti\inst{2},
           D. J. Axon\inst{3,4},
           A. Robinson\inst{3},
            N. Neumayer\inst{5,6}
}
\offprints{A. Gnerucci}

\institute{Dipartimento di Fisica e Astronomia, Universit\`a degli Studi di Firenze, Firenze, Italy\\
             \email{gnerucci@arcetri.astro.it, marconi@arcetri.astro.it}
             \and INAF - Osservatorio Astronomico di Torino, Strada Osservatorio 20, 10025 Pino Torinese, Italy\\
             \email{capetti@oato.inaf.it}
             \and Physics Department, Rochester Institute of Technology, 85 Lomb Memorial Drive, Rochester, NY 14623, USA\\
             \email{djasps@rit.edu, axrsps@rit.edu}
             \and School of Mathematical \& Physical Sciences, University of Sussex, Falmer, Brighton, BN2 9BH, UK
             \and Excellence Cluster Universe, Technische Universit\"at M\"unchen, Boltzmannstr. 2, 85748, Garching bei M\"unchen, Germany\\
             \email{nadine.neumayer@universe-cluster.de}
             \and European Southern Observatory, Karl-Schwarzschild-Str. 2, 85748 Garching bei M\"unchen, Germany\\
             \email{nneumaye@eso.org}
            }

\date{Received ; accepted}
  
 \abstract{We measure the black hole mass in the nearby active galaxy Centaurus A (NGC 5128) using a new method based on spectroastrometry of a rotating gas disk. The spectroastrometric approach consists in measuring the photocenter position of emission lines for different velocity channels. In a previous paper we focused on the basic methodology and the advantages of the spectroastrometric approach with a detailed set of simulations demonstrating the possibilities for black hole mass measurements  going below the conventional spatial resolution.
In this paper we apply the spectroastrometric method to multiple longslit and integral field near infrared spectroscopic observations of Centaurus A. 
We find that the application of the spectroastrometric method provides results perfectly consistent with the more complex classical method based on rotation curves: the measured BH mass is nearly independent of the observational setup and spatial resolution and the spectroastrometric method allows the gas dynamics to be probed down to spatial scales of $\sim 0.02\arcsec$, i.e.~1/10 of the spatial resolution and $\sim 1/50$ of BH sphere of influence radius. The best estimate for the BH mass based on kinematics of the ionised gas is then $\log(M_{BH}\sin i^2/M_\odot ) \simeq 7.5\pm 0.1$ which corresponds to $M_{BH} = 9.6^{+2.5}_{-1.8}\times 10^{7}\,M_\odot$ for an assumed disk inclination of $i=35^\circ$. The complementarity of this method with the classic rotation curve method will allow us to put constraints on the disk inclination which cannot be otherwise derived from spectroastrometry.
With the application to Centaurus A, we have shown that spectroastrometry opens up the possibility of probing spatial scales smaller than the spatial resolution, extending the measured $M_{BH}$ range to new domains which are currently not accessible: smaller BHs in the local universe and similar BHs in more distant galaxies.}

  
   \authorrunning{A. Gnerucci et al.}
     \titlerunning{Spectroastrometry of rotating gas disks: II. Application to Centaurus A.}
     
  \maketitle

\section{Introduction}\label{s1}

One of the fundamental open questions of modern astrophysics is understanding the formation and evolution of the complex structures that characterize the present-day universe such as galaxies and clusters of galaxies. Understanding how galaxies formed and how they become the complex systems we observe today is therefore a major theoretical and observational effort.

There is now strong evidence for the existence of a connection between supermassive black holes (hereafter BHs), nuclear activity and galaxy evolution revealing the so-called co-evolution of black holes and their host galaxies. Such evidence is provided by the discovery of ``relic'' BHs in the center of most nearby galaxies, and by the tight scaling relations between BH masses ($M_{BH}\sim 10^6-10^{10}M_{\astrosun}$) and the structural parameters of the host spheroids like mass, luminosity and stellar velocity dispersion (e.g. \citealt{kormendy-richstone}, \citealt{Gebhardt:2000a}, \citealt{ferrarese-merrit2000}, \citealt{Marconi:2003}, \citealt{Haring:2004}, \citealt{Ferrarese:2005}, \citealt{Graham:2008}). Moreover, while it has long been widely accepted that Active Galactic Nuclei (AGN) are powered by accretion of matter on a supermassive BH, it has recently been possible to show that BH growth is mostly due to accretion of matter during AGN activity, and therefore that most galaxies went through a phase of strong nuclear activity (\citealt{Soltan:1982}, \citealt{Yu:2002a}, \citealt{Marconi:2004}). It is believed that the physical mechanism responsible for this coevolution of BHs an their host galaxies is probably the feedback by the AGN, i.e. the accreting BH, on the host galaxy (\citealt{Silk:1998}, \citealt{Fabian:1999}, \citealt{Granato:2004}, \citealt{Di-Matteo:2005}, \citealt{Menci:2006}, \citealt{Bower:2006}).

The clearest sign of co-evolution, the scaling relations between BH masses and host galaxy properties should be then secured by increasing number, accuracy and mass range of existing measurements. 

Supermassive BHs are detected and their masses measured by studying the kinematics of gas or stars in galaxy nuclei and, currently, there are about $\sim50$ BH mass measurements most of which in the $\sim10^7-10^9M_{\astrosun}$ range (e.g. \citealt{Sani:2010}). The majority of these measurements are made with longslit spectroscopy, but the development in recent years of Integral Field Unit (hereafter IFU) spectrographs has allowed some improvements. IFUs have proven to be powerful tools to study galaxy dynamics as they provide two dimensional coverage of the source without the restrictions of longslit spectrographs, also plagued by unavoidable light losses. Recent studies have presented measurements of BH masses in galactic nuclei using integral field spectroscopy of gas or stellar spectral features (e.g. \citealt{Davies:2006}, \citealt{Nowak:2007}, \citealt{Nowak:2010} ,\citealt{Krajnovic:2007}, \citealt{Krajnovic:2009}, \citealt{Cappellari:2009}, \citealt{Neumayer:2010}, \citealt{Rusli:2010}).
Regardless of the use of longslit or IFU spectrographs, one crucial issue in BH mass measurements is spatial resolution: this must be small enough to spatially resolve the regions where the gravitational effects of the BH can be disentangled from those of the host galaxy. Even with the advent of Adaptive Optics (AO) assisted observations the best spatial resolution achievable are of the order of $\sim 0.1\arcsec$ which corresponds to $\sim 10$ pc at a distance of 20 Mpc.

This paper is the second in a series dealing with gas kinematical BH mass measurements based on a new method.
This method, based on spectroastrometry, provides a simple but accurate way to estimate BH masses and partly overcomes the limitations due to spatial resolution which plague the ``classical'' gas (or stellar) kinematical methods, either using longslit or IFU spectra.
In the first paper of the series (\citealt{Gnerucci:2010}, hereafter Paper I) we illustrated how the technique of spectroastrometry can be used to measure black hole masses focusing on the basis of the spectroastrometric approach and showing with an extended and detailed set of simulations its capabilities and limits. While we mostly focussed on the application of spectroastrometry to longslit spectra, we also showed the technique can be extended to integral field spectra.

In this paper we apply the spectroastrometric method developed in Paper I to estimate the BH mass using real data. 
As a benchmark for our spectroastrometric approach to the study of local BHs, we selected the galaxy Centaurus A (NGC 5128) because it has been extensively studied with the gas kinematical method showing that the gas is circularly rotating and that BH mass and other free parameters are well constrained from the observed kinematics. Moreover both longslit and IFU data are available and this allows a direct comparison of the application of spectroastrometry to different kinds of data.

In Sect.~2 we recap the existing measurements of BH mass for Centaurus A. In Sect.~3 we briefly resume the results of Paper I on the application of spectroastrometry to rotating gas disks for the detection of the central BH. In Sect.~4 we apply the  method to the longslit ISAAC spectra of the nucleus of Centaurus A. In Sect.~5 we apply the method to integral field SINFONI spectra of the nucleus of Centaurus A, both with and without the assistance of Adaptive Optics. Finally, in Sect.~6 we compare and discuss the results from the different datasets, drawing some conclusions on the reliability and accuracy of the method.

\section{Previous measurements of the black hole mass in Centaurus A}\label{s2}

\begin{figure}[!ht]
  \centering
  \includegraphics[width=0.99\linewidth]{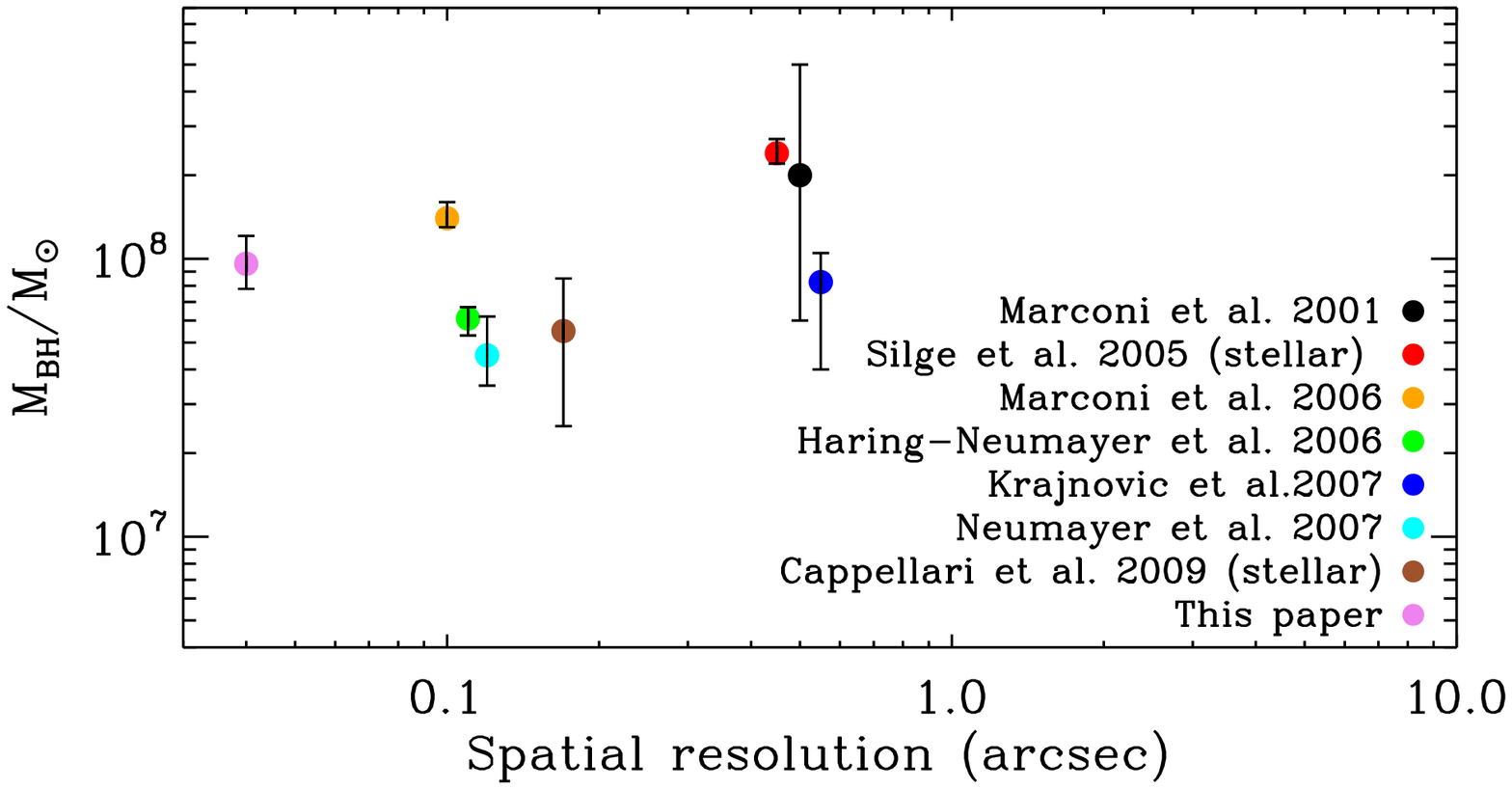}
  \includegraphics[width=0.99\linewidth]{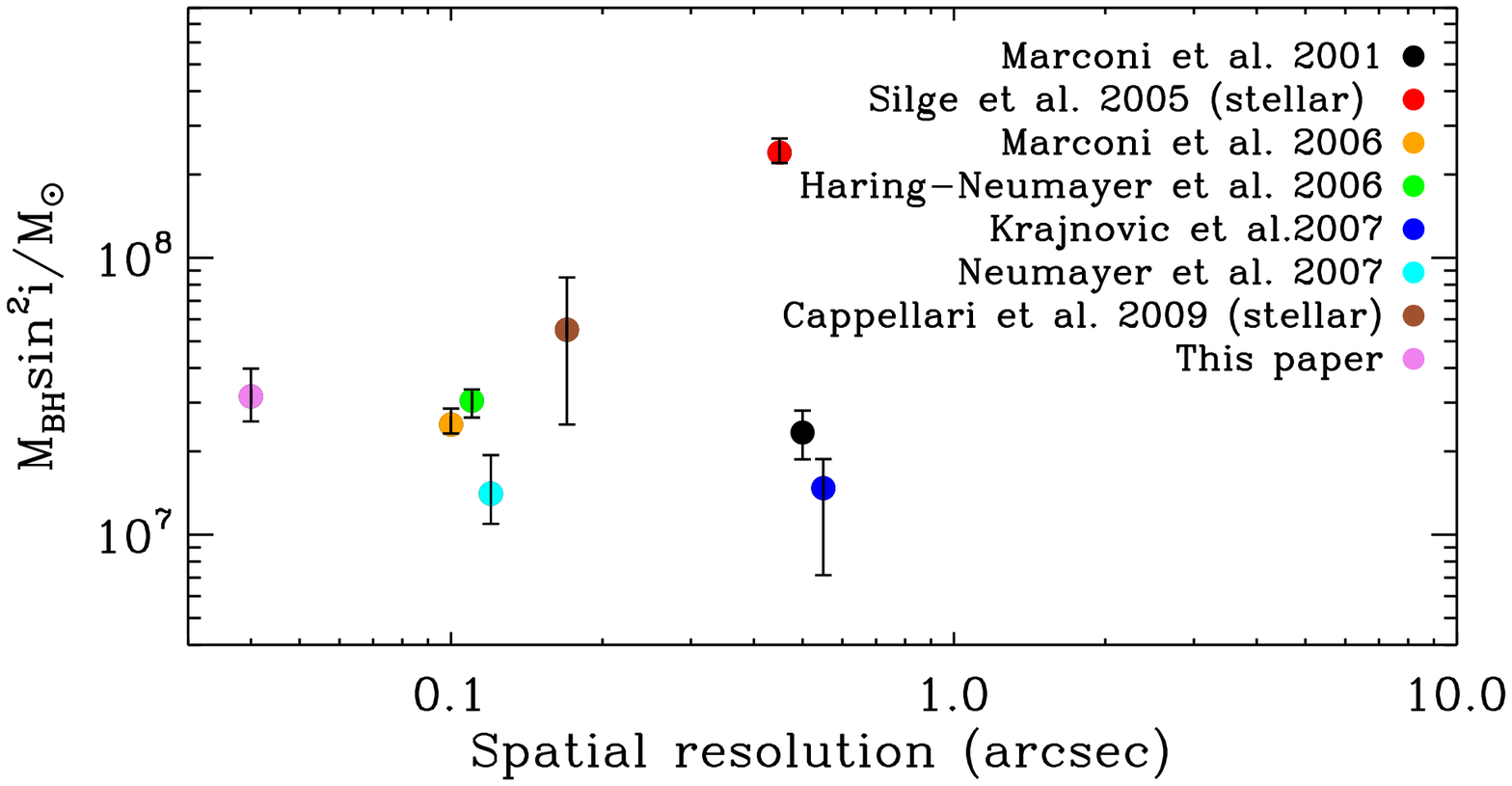}
  \caption{BH mass measurements for Centaurus A from the works mentioned in the text (top panel) and the corresponding $M_{BH}sin^2i$ values (bottom panel) as a function of the spatial resolution of the observations of each measurement (see text). Note that the uncertainties on the measurement by \cite{Marconi:2001} are reduced because they were including uncertainties on $i$.}
     \label{fig01}
  \end{figure}

Existing measurements of the black hole mass in Centaurus A are summarized in Fig.~\ref{fig01} where the various measurements are shown as a function of the spatial resolution of the observations (Full Width at Half Maximum - FWHM - of the Point Spread Function - PSF) used for each measurement. In that Figure we also show the measurement obtained in this paper as a function of the angular resolution actually obtained with the spectroastrometric technique (i.e.~a fraction of the PSF FWHM as will be discussed in detail in the following).

The supermassive black hole in Centaurus A was first detected and its mass measured with a near infrared gas kinematical study using seeing limited spectra obtained with ISAAC at the ESO VLT \citep{Marconi:2001}.
Subsequent higher spatial resolution gas kinematical studies based on longslit spectroscopy were performed using STIS on the HST \citep{Marconi:2006} and AO assisted observations with NAOS-CONICA at the ESO VLT \cite{Haring-Neumayer:2006}.
More recent studies based on integral field spectroscopy were performed by \cite{Krajnovic:2007} using seeing limited observations with CIRPASS at the Gemini South telescope and by \cite{Neumayer:2007}  using AO-assisted observations obtained with SINFONI at VLT.
On the other hand, \cite{Silge:2005} and \cite{Cappellari:2009} performed near infrared stellar kinematical studies based, respectively, on seeing limited longslit spectra (GNIRS at Gemini South) and AO-assisted integral field spectra (SINFONI at the ESO VLT).

The top panel of Fig.~\ref{fig01} shows the different M$_{BH}$ values obtained by the previous authors, spread over almost an order of magnitude. To understand the origin of these differences, in the bottom panel of Fig.~\ref{fig01} we plot the $M_{BH}sin^2i$ values, i.e.~the values constrained by the observed velocity fields and not dependent on the inclination of the rotating gas disks. In the case of the stellar kinematical studies, the authors assumed edge on axisymmetric potentials, therefore no correction is made to obtain $M_{BH}sin^2i$: \cite{Silge:2005} and \cite{Cappellari:2009} discuss the systematic uncertainties associated with their assumptions of edge on models. After removing the inclination effect all gas kinematical measurements show statistical fluctuation within two times the respective sigma; as noted several times, the inclination of the rotating disk is an important source of uncertainty in gas kinematical measurements.

\section{The spectroastrometric measurement of black holes masses}\label{s3}

  \begin{figure*}[!ht]
  \centering
  \includegraphics[width=0.8\linewidth]{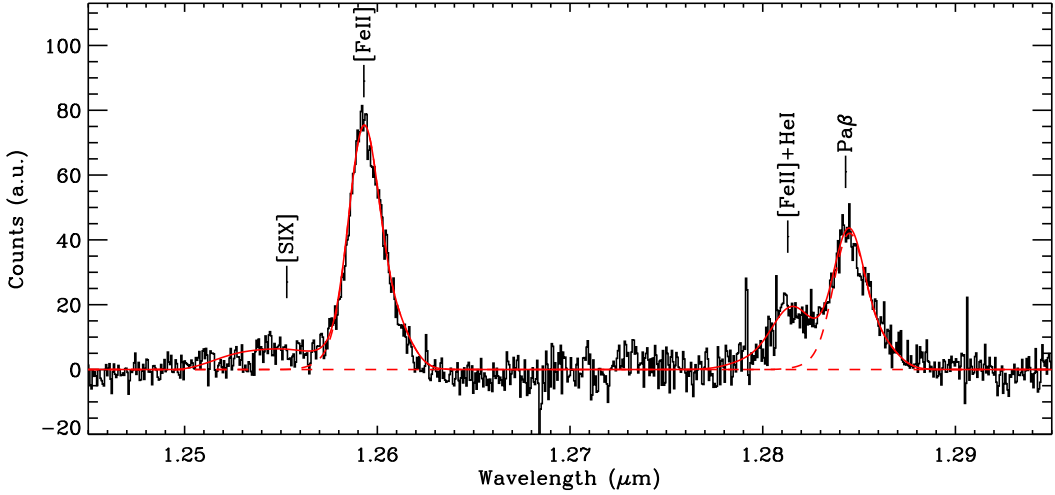}
  \caption{Continuum subtracted ISAAC spectrum (slit PA1) extracted at the position of the continuum peak. Solid red line: simultaneous fit of all four lines. Dashed red lines: the deblended
    Pa$\beta$ and [Fe\,II] components.}
     \label{fig02}
  \end{figure*}
  
In Paper I \citep{Gnerucci:2010} we illustrated how the technique of spectroastrometry can be used to measure the black hole masses at the center of galaxies. In that paper we focused in explaining the basis of the spectroastrometric approach and in showing with an extended and detailed set of simulations how this method is able to probe the principal dynamical parameters of a nuclear gas disk.

The spectroastrometrical method (see \citealt{Bailey:1998}) consists in measuring the photocenter of emission lines in different wavelength or velocity channels. 
It has been used by several authors to study pre-main sequence binaries and the presence of inflows, outflows or the disk structure of the gas surrounding pre-main sequence stars (\citealt{Takami:2003}; \citealt{Baines:2004}; \citealt{Porter:2004,Porter:2005}, \citealt{Whelan:2005}).
We compared this technique with the standard method for gas kinematical studies based on the gas rotation curve and showed that the two methods have complementary approaches to the analysis of spectral data (i.e. the former measures mean positions for given spectrum velocity channels while the latter measures mean velocities for given slit position channels). The principal limit of the rotation curves method resides in the ability to spatially resolve the region where the gravitational potential of the BH dominates with respect to the contribution of the stars. In Paper I we showed that the fundamental advantage of spectroastrometry is its ability to provide information on the galaxy gravitational potential at scales significantly smaller than the spatial resolution of the observations ($\sim1/10$, and better as we will also show in the present work).

The general principle of the spectroastrometric method and its capability in overcoming the spatial resolution limit is illustrated by the following simple example: consider two point-like sources located at a distance significantly smaller than the spatial resolution of the observations; these sources will be seen as spatially unresolved with their relative distance not measurable from a conventional image. However, if spectral features, such as absorption or emission lines at different wavelengths, are present in the spectra of the two sources, the light spatial profiles extracted at these wavelengths will show the two sources separately. From the difference in centroid positions at these two wavelengths one can estimate the separation between the two sources even if this is much smaller that the spatial resolution. This ``overcoming'' of the spatial resolution limit is made possible by the ``spectral'' separation of the two sources.

For clarity, we summarize here the principal features and the main steps of the method presented and discussed extensively in Paper~I.
\begin{itemize}
\item From the longslit spectrum of a continuum-subtracted emission line one constructs the ``spectroastrometric curve'' by measuring the line centroid along the slit for all wavelength channels. The ``spectroastrometric curve'' of the line is given by the position centroids as a function of wavelength.
\item From the simulations presented in Paper~I, we showed that the information about the BH gravitational field is predominantly encoded in the ``high velocity'' (hereafter HV) range of the spectroastrometric curve which comprises the points in the red and blue wings of the line. The HV part of the line spectrum originates from the gas moving at high velocities closer to the BH, in the case of Keplerian rotation and this emission is spatially unresolved. For this reason the HV part of the line spectrum is not strongly influenced by the spatial resolution or other instrumental effects like slit losses or by the intrinsic line flux distribution. On the other hand ``low velocity'' emission (hereafter LV) is usually spatially resolved (i.e.~the gas moving at lower velocities is located farther away from the BH) and this makes the spectroastrometric curve not useful. The HV range of the curve is identified by measuring the spatial extent of the line emission as function of wavelength and discarding the central (i.e.~LV range) bins where the line emission becomes broader than the instrumental spatial resolution.
\item By measuring the spectroastrometric curves of a given line from at least three spectra taken at different slit position angles, one can obtain a ``spectroastrometric map'' of the source on the plane of the sky by geometrically combining the three curves. In the case of integral field spectra, this step is obviously not necessary because the spectroastrometric map is derived directly from the data cubes. In the case of a rotating disk with a radially symmetric line flux distribution the points of the spectroastrometric map should lie on the disk line of nodes. 
However even for IFU data the effect of slit losses or a non-symmetric line flux distribution can perturb the light centroid positions moving them away from the disk line of nodes. As shown in Paper I and discussed above, these effects become negligible for the HV range of the map where the emission is spatially unresolved.  In contrast, the LV points of the map tend to lie away from the line of nodes in a typical ``loop'' shape. The final spectroastrometric map is then obtained by selecting only the HV points.
\item One can then estimate the disk line of nodes by a line fitting of the HV range of the spectroastrometric map, project those points on the estimated disk line of nodes and obtain the disk rotation curve. Finally, one can apply a simple model fitting procedure and obtain the parameters determining the gas rotation curve, and in particular the BH mass.
\end{itemize}


\section{Longslit spectra: observations and data analysis}\label{s4}

\subsection{The data}\label{s41}
We use available near infrared spectra of the nucleus of Centaurus A obtained with ISAAC at the ESO VLT telescope \citep[see][for details]{Marconi:2006}. Briefly, the spectra were obtained with a $0\farcs3$
wide slit and cover a wavelength range of $1.24 - 1.30 \,\mu$m with a resolving power of $\lambda/\Delta\lambda=10500$, corresponding to a spectral
resolution of $\sim1.2\mathrm{\AA}$ at the central wavelength ($\lambda$=1.274 $\mu$m). The spatial scale of the spectra is $0.147\arcsec/$pixel along the slit axis and the dispersion is $0.58\mathrm{\AA}/$pixel. The spatial resolution of the spectra is estimated as $\sim0\farcs5$ (FWHM of the PSF).
There are three different spectra characterized by a different slit position angle: the one we will refer to as ``PA1'' has a position angle of $32.5^{\circ}$, ``PA2'' has $-44.5^{\circ}$ and ``PA3'' has $83.5^{\circ}$ \citep[see][for details]{Marconi:2006}.

\subsection{Analysis of the spectra}\label{s42}

In Fig.~\ref{fig02} we display an example of a typical spectrum of Centaurus~A extracted from one pixel along one of the slits: in particular, this
corresponds to the ``PA1'' continuum subtracted spectrum, extracted at the position of the continuum peak. Several gas emission lines can be identified:
Pa$\beta$ at $\sim1.284 \mu$m, [Fe\,II] and He\,I at $\sim1.281 \mu$m, [Fe\,II] at $\sim1.259 \mu$m and [S\,IX] at $\sim1.255 \mu$m. The lines used
for the gas kinematics by \cite{Marconi:2006} are the Pa$\beta$ and the [Fe\,II] lines and we will also concentrate on these.

  \begin{figure}[!h]
  \centering
  \includegraphics[width=\linewidth, trim=10 0 0 0]{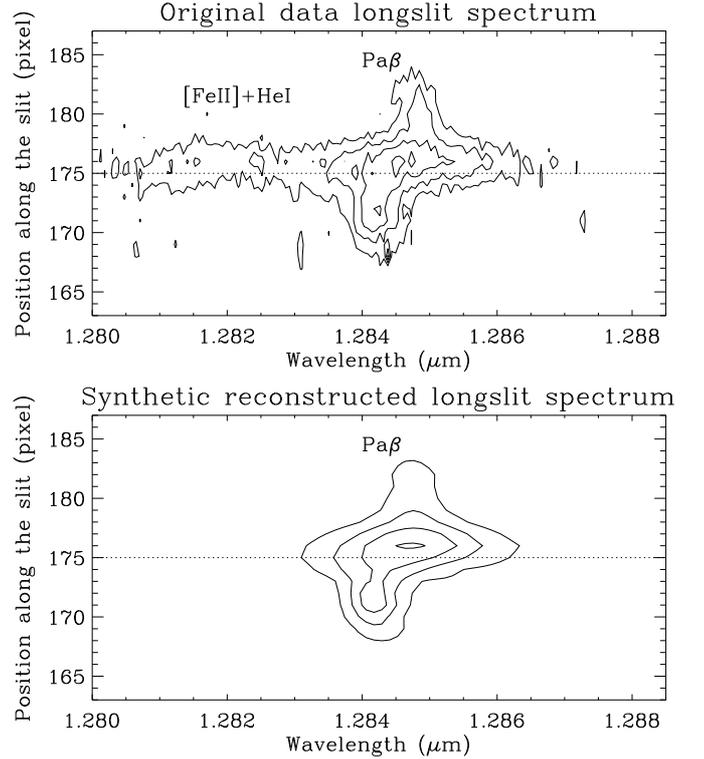}
  \caption{Position velocity diagram of the observed ISAAC spectrum at PA1 near to the Pa$\beta$ line. Upper panel: observed spectrum of the Pa$\beta$ and [Fe\,II]+He\,I complex. Bottom panel: ``synthetic'' reconstruction of the deblended Pa$\beta$ spectrum. The horizontal dotted line overplotted on each panel represents the continuum peak position. The isophotes denote the same values in both panels.}
        \label{fig03}
  \end{figure}

  \begin{figure*}[!ht]
  \centering
  \includegraphics[width=\linewidth]{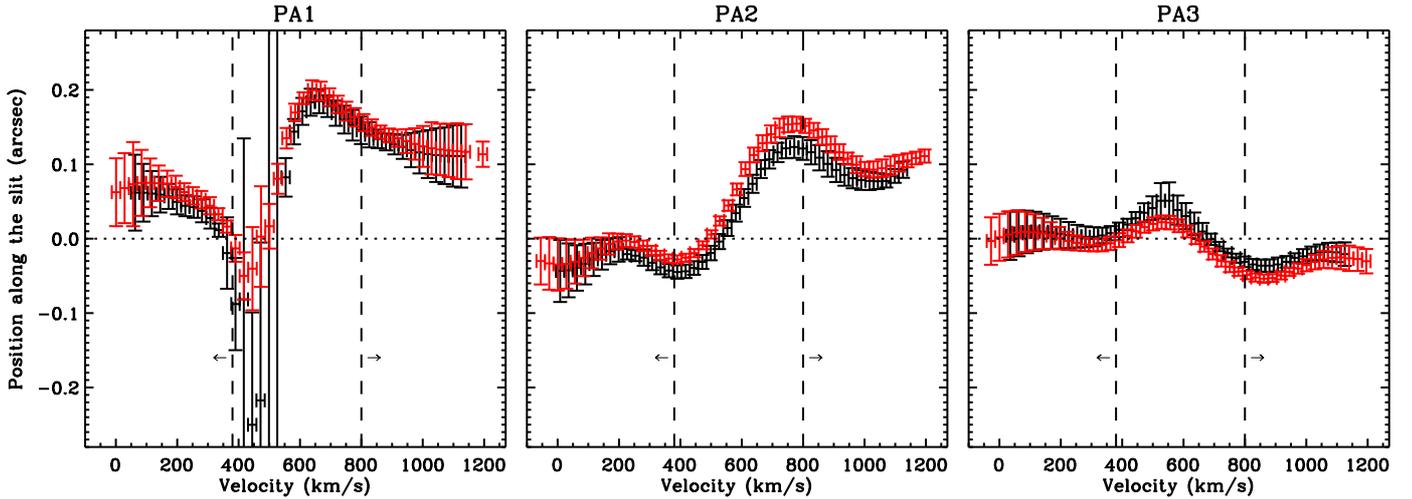}
     \caption{Spectroastrometric curves of the Pa$\beta$ (black points) and [Fe\,II] (red points) lines (ISAAC spectra) at the three slit position angles. Left panel: PA1. Central panel: PA2. Right panel: PA3. The dashed vertical lines on each panel represent the limits of the HV range.}
        \label{fig04}
  \end{figure*}

  As clearly visible in the figure, both the Pa$\beta$ and [Fe\,II] lines  are blended on the
  blue sides with [Fe\,II]+He\,I and [S\,IX], respectively. Potentially this constitutes a problem in measuring the spectroastrometric
  curve because the blended lines will certainly affect the position of the light centroid along the slit for a given blue velocity and the position
  centroid will not be indicative anymore of the mean position of the gas at a given line of sight velocity.
  
  In order to solve this problem it is necessary to deblend the lines under examination and therefore we performed a simultaneous fit of the 4 lines,
  each with a Gauss-Hermite function (the red solid line), at all slit positions along the PA1, PA2 and PA3 slits.  Following
  \citet{Marconi:2006} we assume that all lines, hence Pa$\beta$ and [Fe\,II], share the same kinematics: in the simultaneous fit all lines are
  constrained to the same velocity, velocity dispersion and Hermite parameters ($h_3$ and $h_4$), while they can have different line fluxes.
  Fig. \ref{fig02} shows an example of such a fit.

  To obtain the spectroastrometric curve we considered all fitted profiles of Pa$\beta$ and [Fe\,II] and we reconstructed synthetic longslit spectra of the
  emission lines, cleaned in terms of noise and blended lines. In Fig. \ref{fig03} we show an example of one original Pa$\beta$ longslit spectrum
  compared with its ``synthetic'' version where one can notice that the noise has been smoothed away and the [Fe\,II]+HeI complex has been removed. It
  should be noticed that whereas we constrained Pa$\beta$ and [Fe\,II] to the same kinematics, their fluxes can be different, therefore, in terms of the
  spectroastrometric analysis, the two lines are distinct and can still provide different results. Additionally the ``synthetic'' reconstructed spectrum is
  noise-free because each row represents the fitted parametric profile but one has to take into account the errors on the free parameters in order to
  estimate the errors on each synthetic pixel counts. Therefore, for each spectral fit, we simulated $1000$ synthetic spectra from 1000 realization of the set of
  the five parameters distributed following a pentavariate distribution with the fit correlation matrix. The flux and error of each pixel is then
  estimated from mean and standard deviation of the 1000 realizations.  We note that for each profile along the slit, the errors on fluxes are uncorrelated because they originate from independent fits to the line profiles.

  From the ``synthetic'' Pa$\beta$ and [Fe\,II] spectra for the three slits ($PA1$, $PA2$ and $PA3$), following the method outlined above, one can derive the spectroastrometric curves which are shown in Fig. \ref{fig04}. Wavelengths are converted in velocity using as reference (zero velocity) the rest frame Pa$\beta$ and [Fe\,II] wavelengths (respectively $1.28216\mu m$ and $1.25702\mu m$).

  \begin{figure*}[!ht]
  \centering
  \includegraphics[width=0.9\linewidth, trim=10 5 10 20]{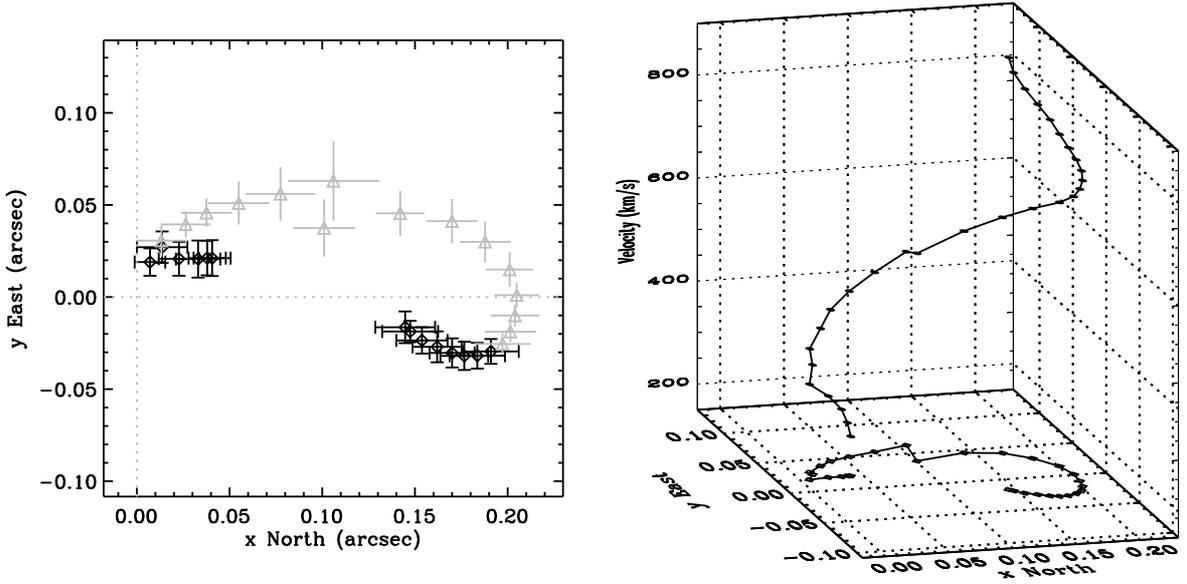}
  \caption{Spectroastrometric 2D map derived from the Pa$\beta$ line (ISAAC spectra). Left panel: derived photocenter positions on the sky plane, the black points
    are those actually used for the minimization. Right panel: the 3D plot of the map, where the z axis is velocity.}
        \label{fig05}
  \end{figure*}
  
Fig. \ref{fig04} reveals features which needs further comments. The extremely large error bars observed in the Pa$\beta$ curve at PA1
($400km/s\lesssim v \lesssim 550km/s$) are due to the fact that the light profile is spatially resolved and in this range our method of measuring the centroid position
is not reliable. As observed in section 2.3 this is expected in the LV range, but the relevant information are concentrated in the HV range of the curve.
Except for those points, errors on photocenter positions range from $\sim 0.01$ to $\sim0.05\arcsec$ that is from $\sim 1/50$ to $\sim 1/10$ of the spatial resolution of the data and this is the accuracy with which we can measure centroid positions.

As expected, the spectroastrometric curves for the two lines are marginally different. Their differences are due to the intrinsic flux distribution of the lines on the sky plane but, as observed in section 2.3.4, these differences tend to disappear in the HV range.

\subsection{The spectroastrometric map of the source}\label{s43}
\label{map}
For each emission line we have obtained three spectroastrometric curves, one for each PA of the slit.  Each spectroastrometric curve provides the
photocenter position along one slit, i.e. the position of the photocenter projected along the axis defined by the slit direction. Combining the
spectroastrometric curves we can thus obtain the map of photocenter positions on the plane of the sky for each velocity bin. In principle, the
spectroastrometric curves from two non-parallel slits should suffice but we can use the redundant information from the three slits to recover the 2D sky
map as described in Section 4.1 of Paper I. We have chosen a reference frame in the plane of the sky centered on the center of PA1 slit (that correspond to the position of the continuum peak along the slit) with the X axis along the North direction. For a given velocity bin we then determined the position of the light centroid on the sky plane resulting in the 2D spectroastrometric map shown in Fig. \ref{fig05}.  

Note that the coordinates on the plane of the sky of the center of the PA2 and PA3 slits must be considered as free parameters. These unknowns are estimated simultaneously with the position of the photocenter following a $\chi^2$ minimization procedure. The final spectroastrometric map on the plane of the sky is that given by the best
fitting set of slit centers.  The error bars on the points represent the uncertainties resulting from the fit. The black points correspond to the HV range (i.e. $v\lesssim380$ km$/s$ and $v\gtrsim800$ km$/s$) which were actually used to determine the location of slit centers. Indeed, in Paper I we concluded that the HV range of the spectroastrometric curve is more robust, and less affected by slit losses which artificially change the photocenter position in the LV range.

\begin{figure}[!h]
  \centering
  \includegraphics[width=\linewidth]{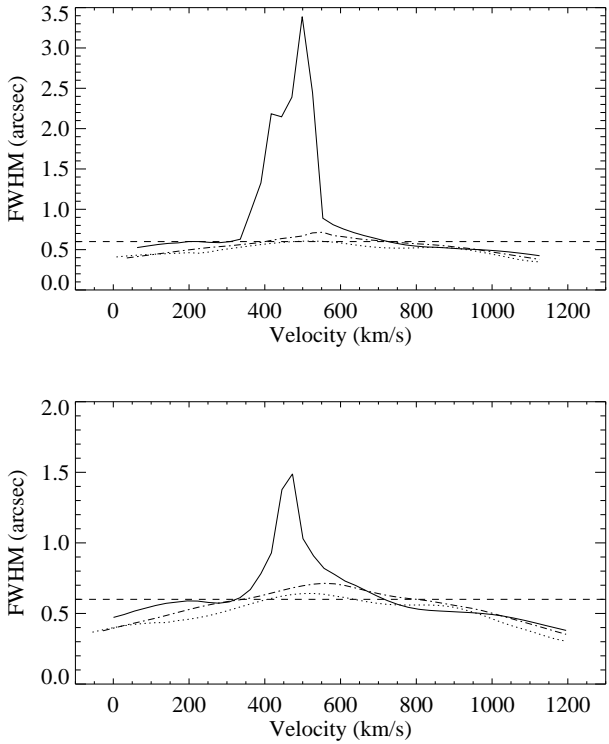}
     \caption{FWHM of the Gaussian fitted to the principal peak of the light profile for the ISAAC emission line spectra. Upper panel: Pa$\beta$ line. Bottom panel: $[FeII]$ line. Solid lines: PA1 curves. Dotted lines: PA2 curves. Dot-dashed lines: PA3 curves. The horizontal dashed lines denotes $1.1$ times the PSF FWHM.}
        \label{fig06}
  \end{figure}

As observed in appendix A of Paper I, we used the width  of the Gaussian fitted to the principal peak of the light profile to select the HV range: in Fig. \ref{fig06} we display the FWHMs (Full Width Half Maxima) of the Gaussian fitted to the principal peak of the light profile for spectroastrometric curves of Centaurus A.
 We can observe that in the LV range the FWHM increases because the emission peak is spatially resolved. We compare the FWHM to the spatial resolution ($\sim0.5\arcsec$) because the FWHM of the light profile of an unresolved source should be of the order of the spatial resolution. We selected the HV range by imposing that the FWHM is lower than $1.1$ times the spatial resolution (FWHM of the PSF), resulting in  $v\lesssim380$km$/s$ and $v\gtrsim800$km$/s$ for the spectroastrometric curves of both lines.

  The two-dimensional spectroastrometric map just described (and shown in Fig.~\ref{fig05}) can now be used to estimate some geometrical parameters of the nuclear gas
disk. If the gas kinematics are dominated by rotation around a point-like mass (the BH), the position of the light centroid in the HV range should
lie on a straight line (which identifies the direction of the disk line of nodes) and should approach, at increasing velocities, the position of the BH.
These considerations allows us to estimate the position angle of the line of nodes ($\theta_{LON}$) by fitting a straight line to the HV points, considering their errors in both X and Y directions (see Fig. \ref{fig05}). The results of these fits are reported in Table \ref{tab1} together with the formal fit uncertainties. We verified the reliability of these uncertainties using the bootstrap method \citep{efron:1994}. In particular, we randomly extracted 100 datasets from the HV points and re-performed the fitting of the line of nodes. Due to the random extraction, the new datasets will have the same number of points as the original one but with some points replicated a few times and others entirely missing; we thus randomly assign different weights in the fits  of the HV point. After performing the 100 fits, we estimate the error on $\theta_{LON}$ by taking the standard deviation of the best fit values which are usually normally distributed. This error is consistent with the formal fit error.

 We can also make a first estimate of the BH location by taking the average position of the HV points in the spectroastrometric map. These positions are then refined with the model fitting procedure described below and are reported in Table \ref{tab1}. Estimates from different lines are all consistent with each other when taking into account  the  $\sim0.01\arcsec$ uncertainties (~$\sim1/20$ of the spatial resolution of the data).
 
In top panels of Fig.~\ref{fig10} we show the derived spectroastrometric maps for the Pa$\beta$ band [Fe\,II] lines. It should be noticed that all LV points lie outside of the line of nodes, as expected. 

  \begin{figure*}[ht!]
  \centering
  \includegraphics[width=0.48\linewidth, trim=0 10 0 33]{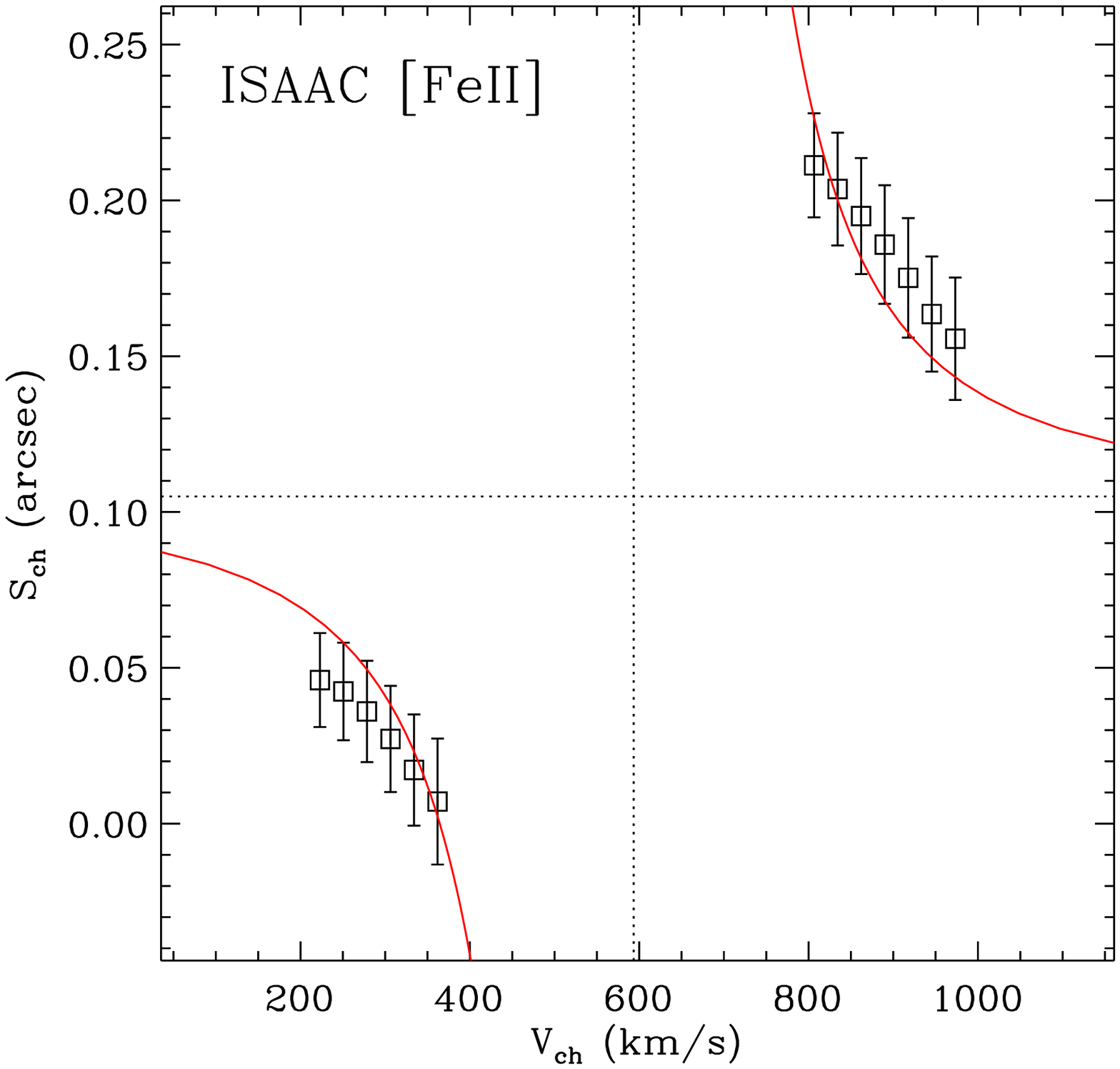}
  \includegraphics[width=0.48\linewidth, trim=0 10 0 33]{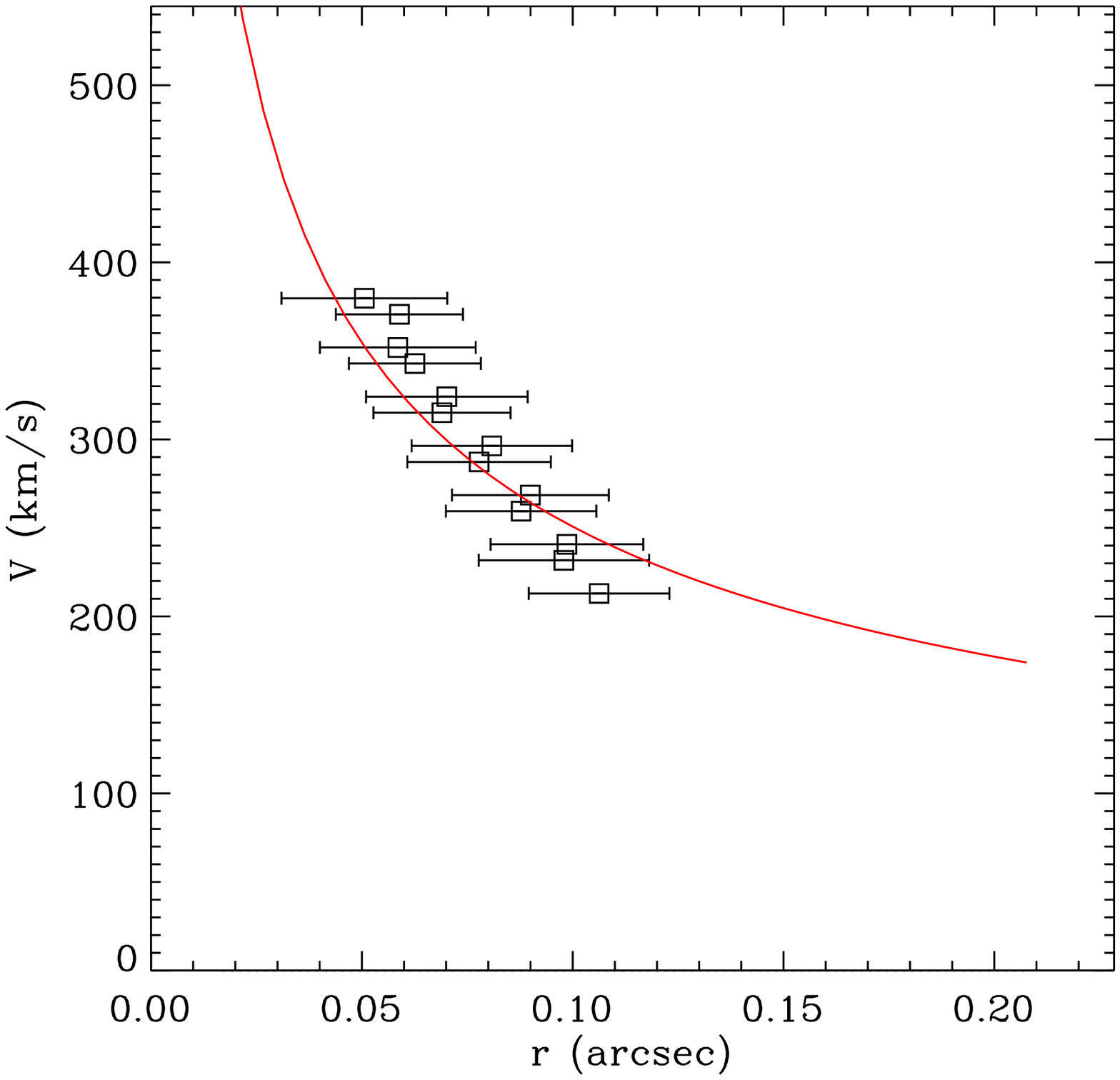}
  \includegraphics[width=0.48\linewidth, trim=0 10 0 20]{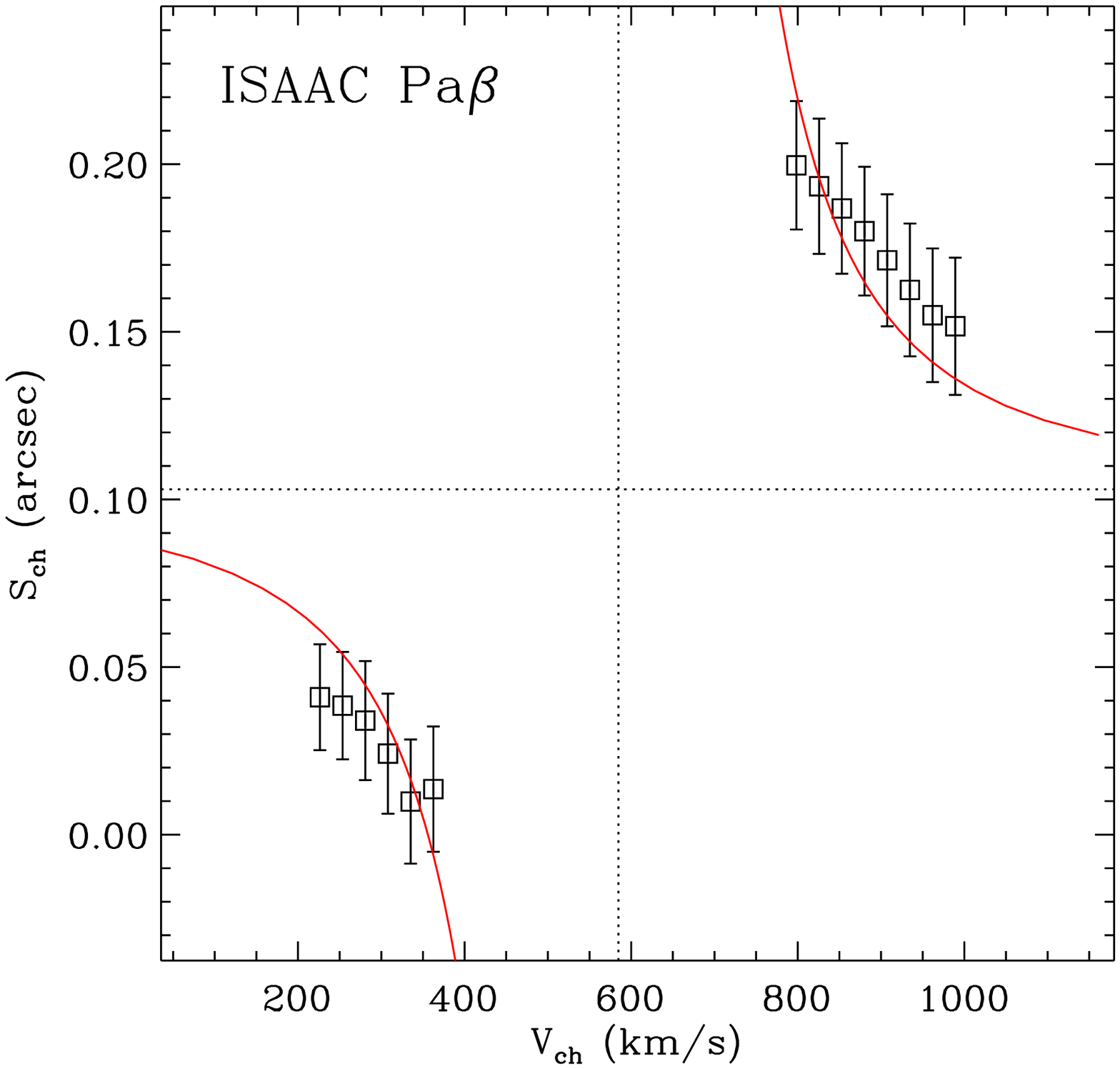}
  \includegraphics[width=0.48\linewidth, trim=0 10 0 20]{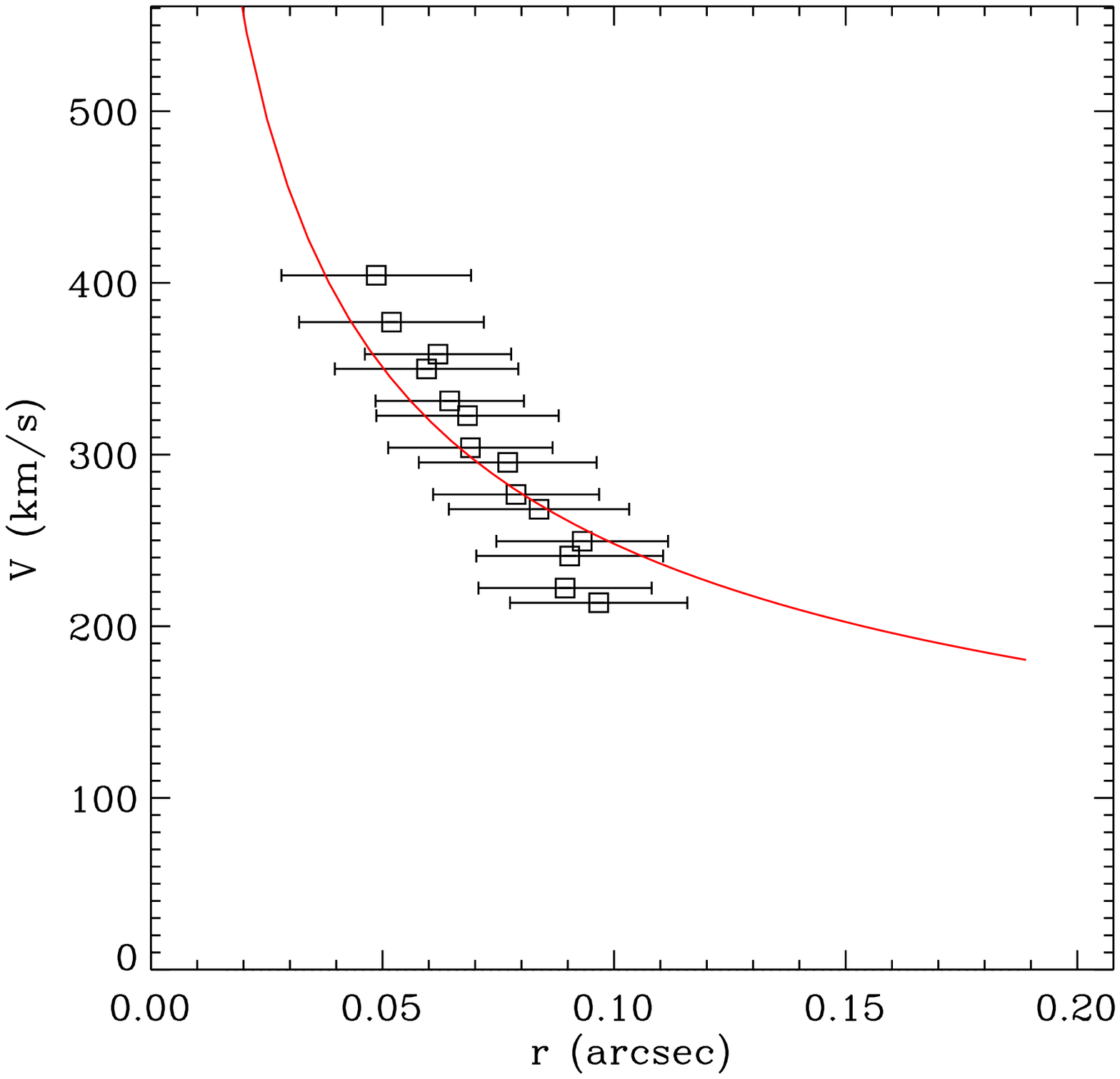}
  \includegraphics[width=0.48\linewidth, trim=0 10 0 20]{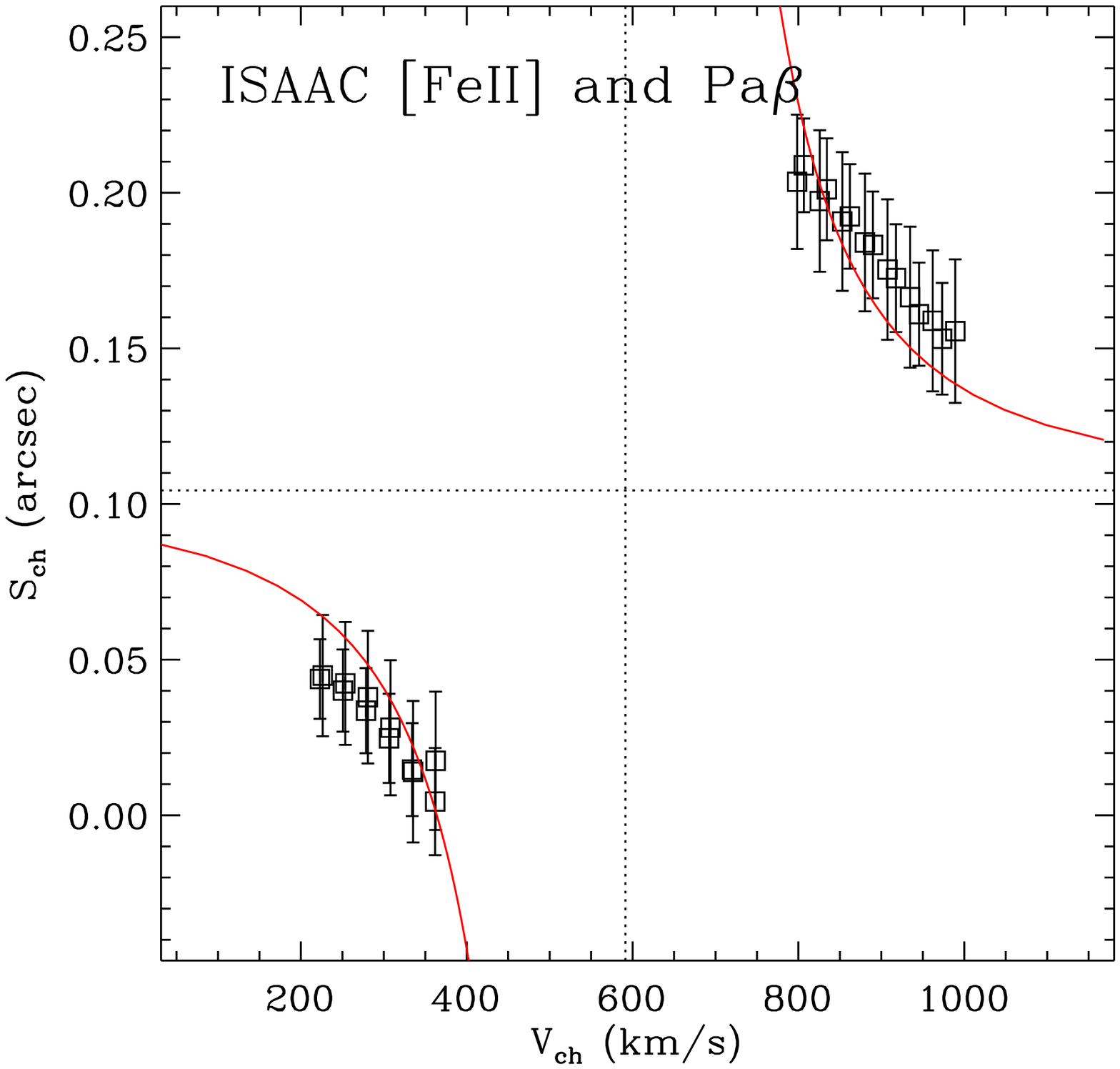}
  \includegraphics[width=0.48\linewidth, trim=0 10 0 20]{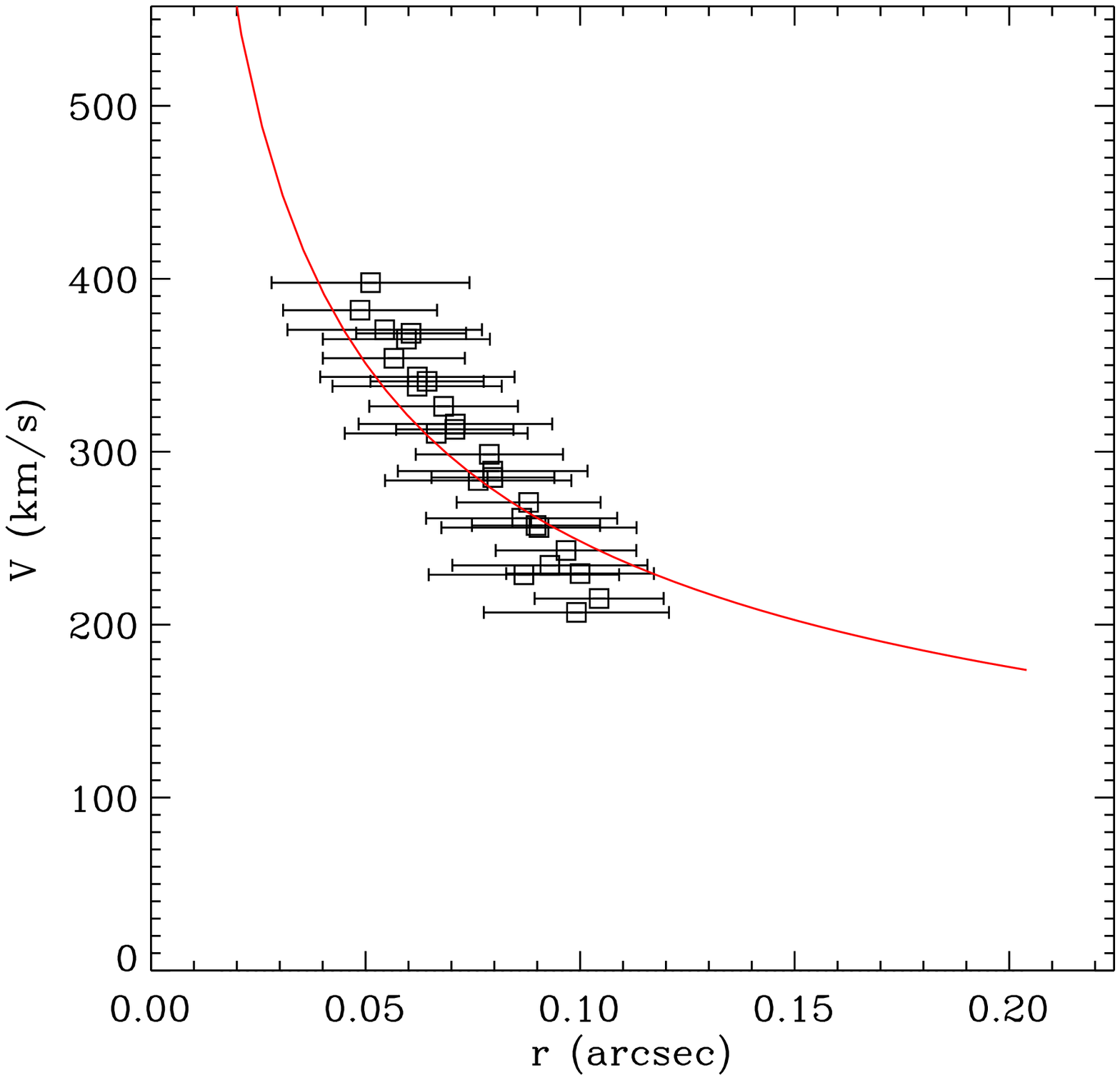}
     \caption{Results of the spectroastrometric modeling of the ISAAC [Fe II] data (upper panels), ISAAC Pa$\beta$ data (central panels) and simultaneous fit of the $[FeII]$ and Pa$\beta$ data (lower panels). The left panels show the line of nodes projected rotation curve $S_{ch}$ vs. $V_{chan}$. The right panels show the $r=|S_{ch}-S_0|$ vs. $V=|V_{ch}-V_{sys}|$ rotation curve. The solid red lines represent the curves expected from the model}
        \label{fig08}
  \end{figure*}

\subsection{Estimate of the BH mass from the spectroastrometric map}\label{s44}

Here we recover the BH mass value from the spectroastrometric map, following the method outlined in Sect.~5 of Paper I.

Briefly, under the assumption that the gas lies in a thin disk configuration inclined by $i$ with respect to the plane of the sky ($i=0$ face-on) and the disk line of nodes has a position angle $\theta_{LON}$, the circular velocity of a gas particle with distance $r$ from the BH is given by:

\begin{equation}
V_{rot}=\sqrt{\frac{G[M_{BH}+M/L\cdot L(r)]}{r}}
\label{1}
\end{equation}

where $r$ is the distance to the BH, $L(r)$ is the radial luminosity density distribution in the galactic
nucleus and $M/L$ is the mass to light ratio of the stars (see Paper I for more details).

The component  of $V_{rot}$ along the line of sight (hereafter $V_{ch}$ for ``channel velocity'') is:
\begin{equation}
\bar{V}_{ch}=V_{rot}\sin(i)+V_{sys}
\label{2}
\end{equation}
where $i$ is the inclination of the disk and we also added the systemic velocity of the galaxy $V_{sys}$ \citep[see][for details]{Marconi:2006}.

As explained in section \ref{s42} and \ref{s43} and sections 3 and 5 of Paper I, we only  make use of the HV points of the spectroastrometric 2d map to estimate the BH mass.

The first step is to recover the disk line of nodes by fitting a straight line on the 2d map. We then project the position of the 2d map points  $(x_{ch}, y_{ch})$ on the line of nodes, calculating their coordinate with respect to this reference axis  ($S_{ch}$) and then their distance $r$ from the BH used in Eq. \ref{1} (i.e. $r=k|S_{ch}-S_0|$ where $S_0$ is the coordinate along the line of nodes of the BH and $k$ is a scale factor to transform arcsec in the right distance unit\footnote{To be consistent with all previous measurements we assume a distance to Centaurus A of $3.5 Mpc$. At this distance $1\arcsec$ corresponds to $\sim17pc$.}; see Paper I for details).

Then from Eqs. \ref{1} and \ref{2} we obtain the model channel velocity:
\begin{equation}
\bar{V}_{ch}=\pm\sqrt{\frac{G(M_{BH}+M/L\cdot L(k|S_{ch}-S_0|)}{k|S_{ch}-S_0|}}sin(i)+V_{sys}
\label{3}
\end{equation}

and the sign depends on the side of the disk considered. The unknown parameters of this model are found minimizing the quantity

\begin{equation}
\chi^2=\sum_{ch}{\left[\frac{V_{ch}-\bar{V}_{ch}}{\Delta(S_{ch}; par)}\right]^2}
\label{4}
\end{equation}

where $\Delta(S_{ch}; par)$ is the uncertainty  of the numerator. As previously discussed, we restrict the fit (i.e. the sum over the velocity channels) to the HV range. As explained in sect.~5 of Paper I, the channel velocity $V_{chan}$ has no associated uncertainty since it is not a measured quantity but the central value of the velocity bin. Finally we add a constant error ($\Delta_{sys}$) in quadrature to the quantity $\Delta(S_{ch}; par)$ in order to obtain a reduced $\chi^2$ close to $1$ (see sect.~5 of Paper I for a detailed explanation of this choice).
Finally, using the best fit values of the model parameters we can compute the $(x,y)$ position of the BH in the sky plane.

We verified that the BH mass value is insensitive to the actual value of the position angle of the line of nodes. Indeed, when repeating the fit with $\theta_{LON}$ values varying within the uncertainties, $M_{BH}$ changes by only $\sim 0.02$dex, well below the typical $1\sigma$ uncertainties of the fit (see Table \ref{tab1}).
We have also checked if the final BH mass estimate could be biased by incorrect estimates of both $\theta_{LON}$  and the BH position. To do this, we repeated the analysis including only the HV points which deviate by more than $2\sigma$ from the linear fit to the line of nodes. The final BH mass changes by less than 0.1dex.

In the rotation curve model shown in Eq.~\ref{3}, disk inclination $i$ and BH mass are coupled since coordinates along the line of nodes ($S_{ch}$) do not depend on $i$ and this parameter appears only as a scaling factor on the velocity. The reason for this coupling is that we are effectively measuring velocities of rotating material located on the line of nodes, thus removing any dependence on $i$ except for the projection of the velocity along the line of sight. Therefore our fitting method can only measure $M_{BH}\,\sin^2i$, and we need to assume a value for the inclination to obtain a value of the mass.
In conclusion the free parameters in our fit are:
\[
  \begin{array}{lp{0.8\linewidth}}
     M_{BH}\,\sin^2i       & mass of the BH;     \\
     M/L\,\sin^2i         & mass to light ratio of the nuclear stars;\\
     S_0      & line of nodes coordinate of the BH;\\
     V_{sys}      & systemic velocity of the galaxy;\\
  \end{array}
\]
in the following we will only report $M\,sin^2i$ values and we will discuss the inclination values we assume and the relative mass values we obtain.

We have performed the fit of the spectroastrometric data from the $[FeII]$ and
Pa$\beta$ lines, both separately and simultaneously. Fit results are tabulated in Table \ref{tab1} and presented graphically in Fig. \ref{fig08} where we plot the $r=|S_{ch}-S_0|$ vs. $V=|V_{ch}-V_{sys}|$ rotation curve and the line of nodes projected rotation curve (e.g.  $S_{ch}$ vs. $V_{ch}$). The solid red lines represent the curves expected from the model  ($r$ vs. $|\bar{V}_{ch}-V_{sys}|$ and $S_{ch}$ vs. $\bar{V}_{ch}$ respectively).

The first result is that assuming a disk inclination of $25^{\circ}$ as in \citet{Marconi:2006} we obtain $M_{BH}=10^{8.14\pm0.02}M_{\astrosun}$, perfectly consistent with the result presented in that work (\citealt{Marconi:2006} report $M_{BH}=10^{8.14\pm0.04}M_{\astrosun}$). Since we are actually using the same ISAAC data of \citealt{Marconi:2006}, we can draw the important conclusion that the spectroastrometric method provides results perfectly consistent with the rotation curve method, but with a much simpler approach which does not require to take into account complex instrumental effects.

Furthermore, we note that $M/L$ is not constrained, a result which has been already found with the classical rotation curve fitting. 
Computing the radius of the BH sphere of influence  for Centaurus A and using a stellar velocity dispersion of $\sigma_{star}\simeq150km/s$ \citep{Gultekin:2009} one obtains $r_{BH}\simeq25pc$, a factor of $20$ larger than the distances we are considering here (at the distance of $3.5$Mpc $1\arcsec$ corresponds to $\sim17pc$ and so the apparent dimension of $r_{BH}$ is $\sim1.5\arcsec$). This implies that at these small radii the contribution of the stellar mass to the gravitational potential is negligible and consequently $M/L$ cannot be constrained with the fit. 

Another important result derived from the application of the spectroastrometric method is that the minimum distance from the BH at which there is a velocity estimate 
is $\sim0.05\arcsec$ corresponding to $\sim0.8$ pc, while with the standard rotation curve method the minimum distance from the BH which can 
be observed is of the order of half the spatial resolution ($0.25\arcsec$). This clearly shows how spectroastrometry can overcome the spatial resolution 
limit. In Centaurus A the sphere of influence of the BH is already well resolved in ground based observations  with good seeing and this enables us to obtain a BH
measurement already from the standard methods. The potential of spectroastrometry clearly relies on the possibility of using the extra
resolution to measure the masses of BHs with smaller spheres of influence, like those with lower masses, or located in more distant galaxies.
  \begin{figure}[!h]
  \centering
  \includegraphics[width=\linewidth]{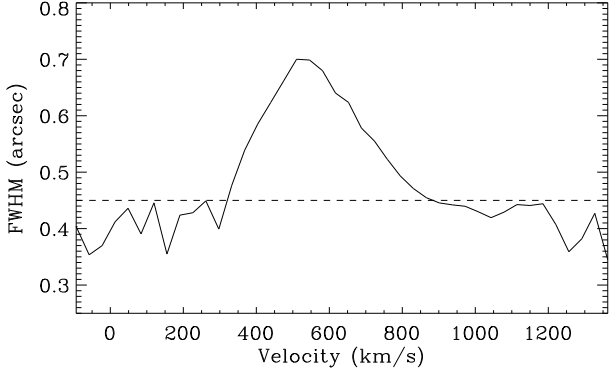}
     \caption{Example of the FWHM of the fitted 2D gaussian to each velocity channel map for the H band seeing limited SINFONI [FeII] data. The horizontal dashed lines denotes $1.1$ times the PSF FWHM.}
        \label{fig09}
  \end{figure}
  \begin{figure*}[!ht]
  \centering
  \includegraphics[width=0.49\linewidth]{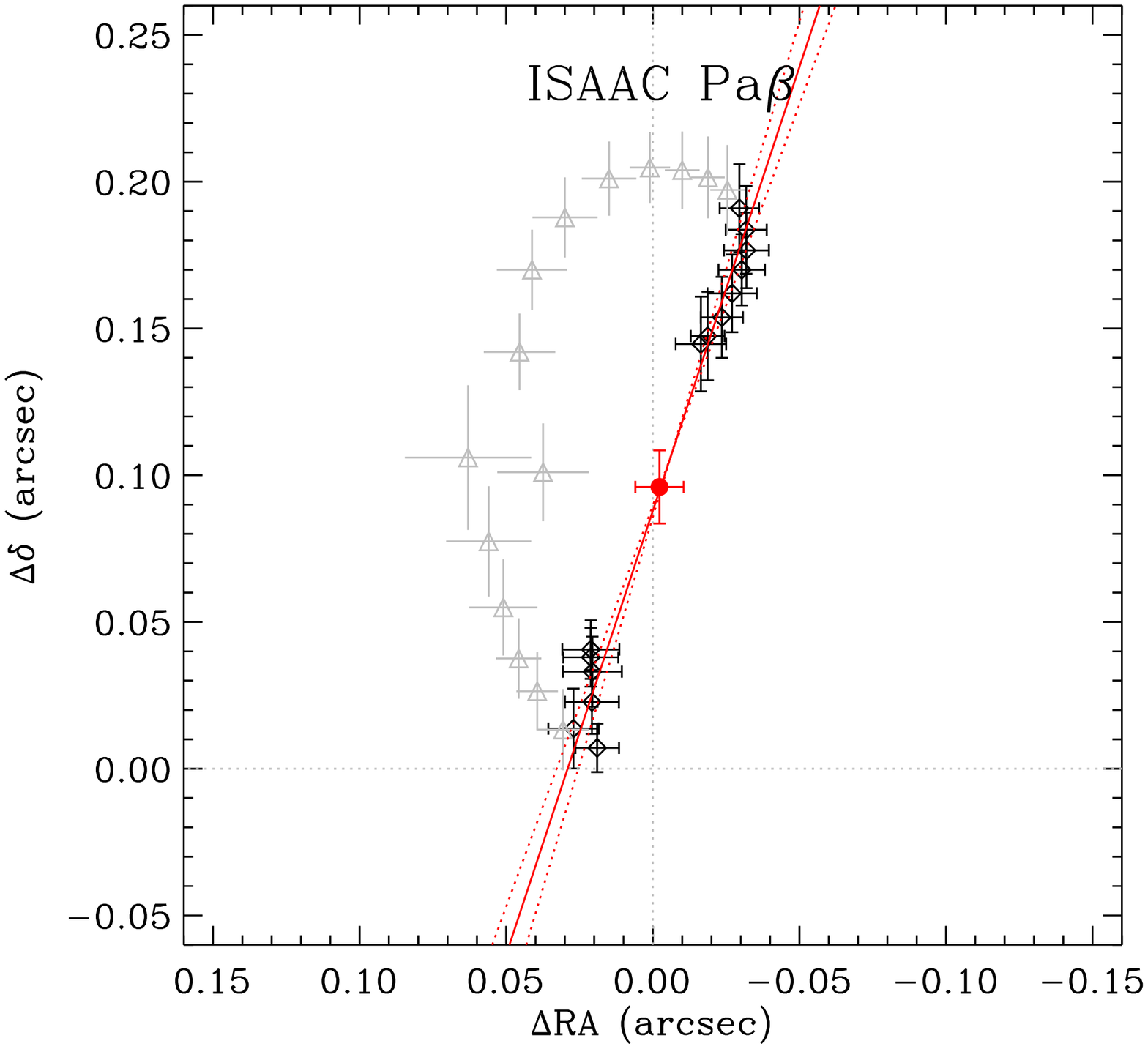}
  \includegraphics[width=0.49\linewidth]{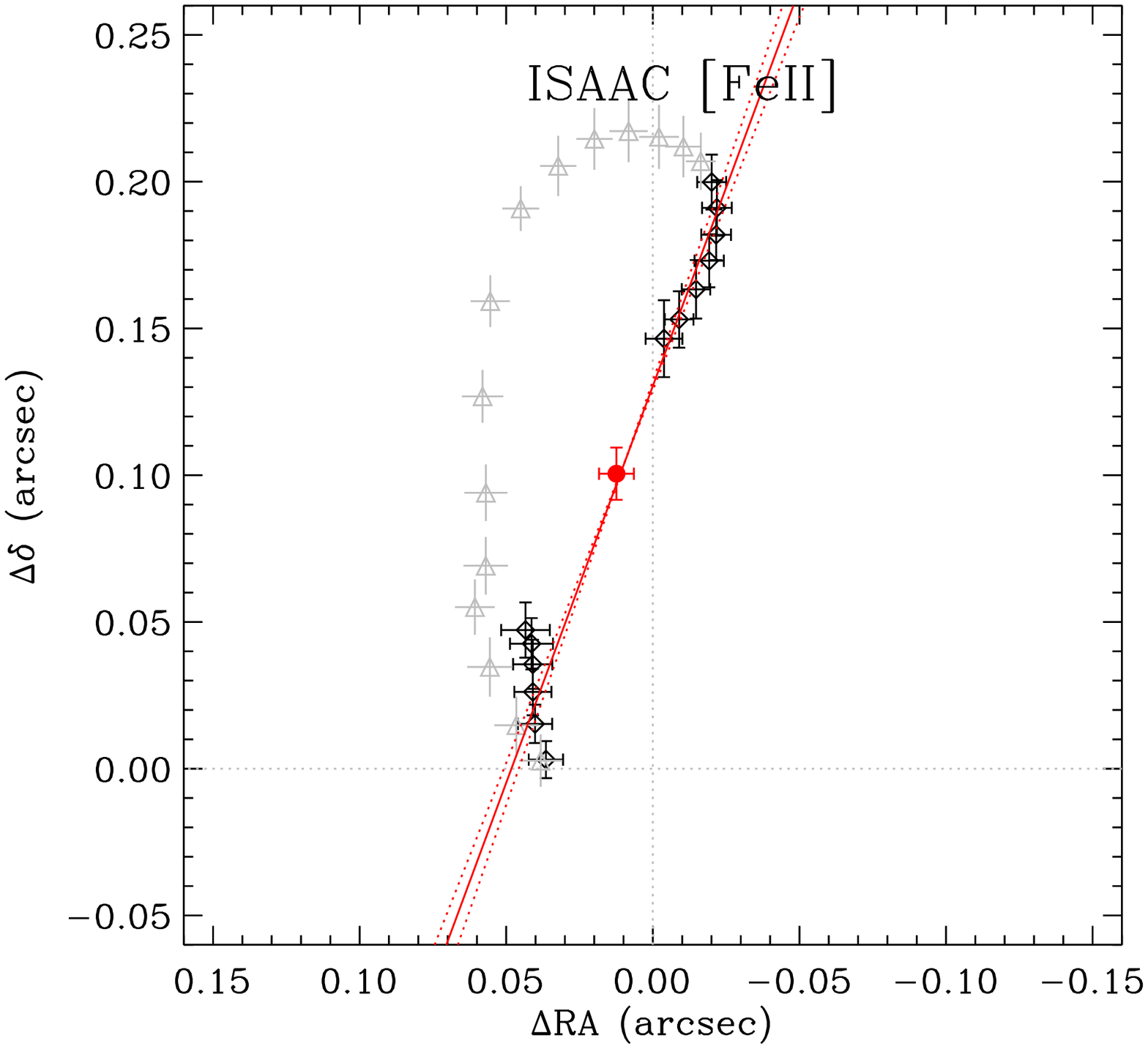}
  \includegraphics[width=0.49\linewidth]{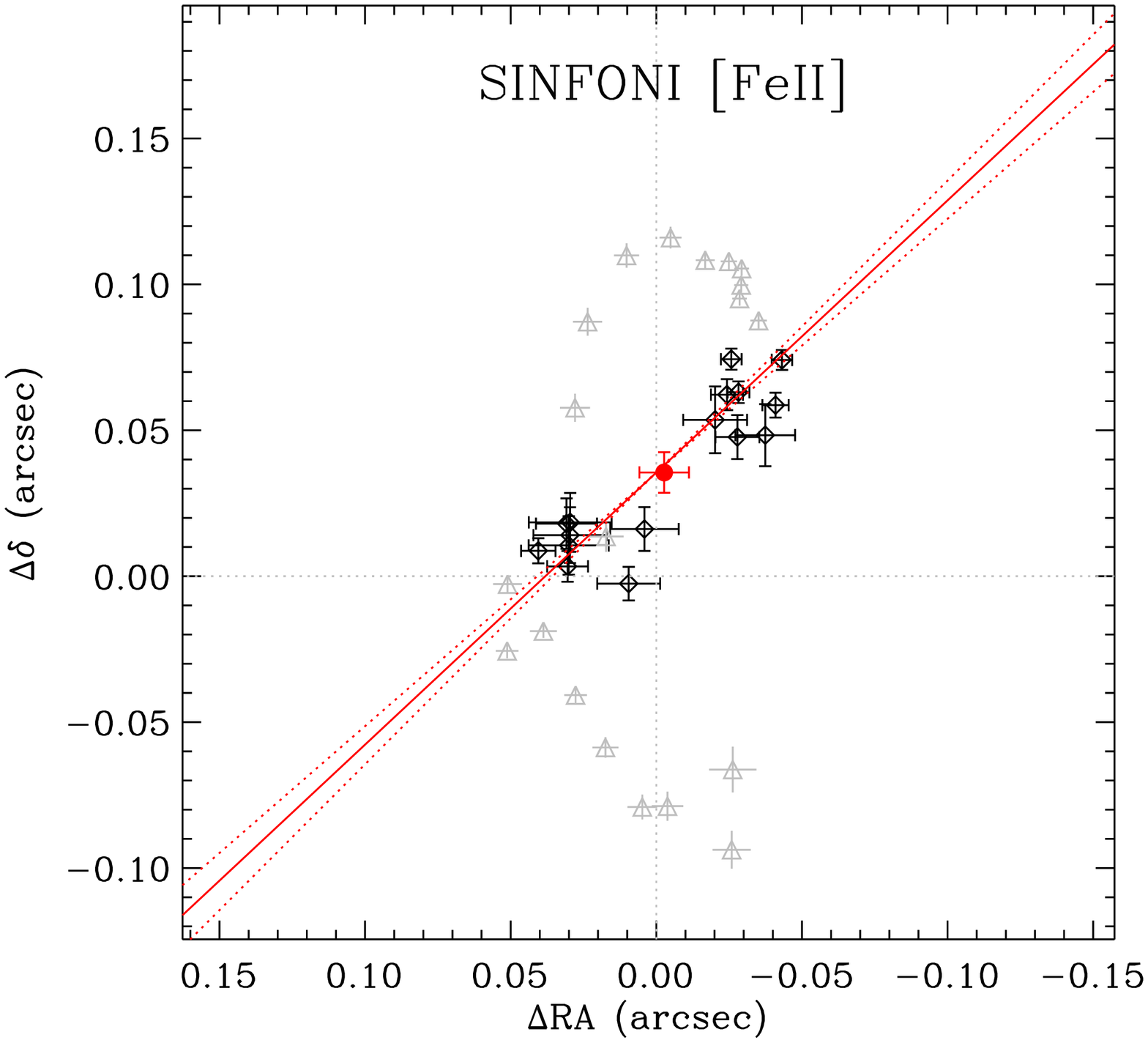}
  \includegraphics[width=0.49\linewidth]{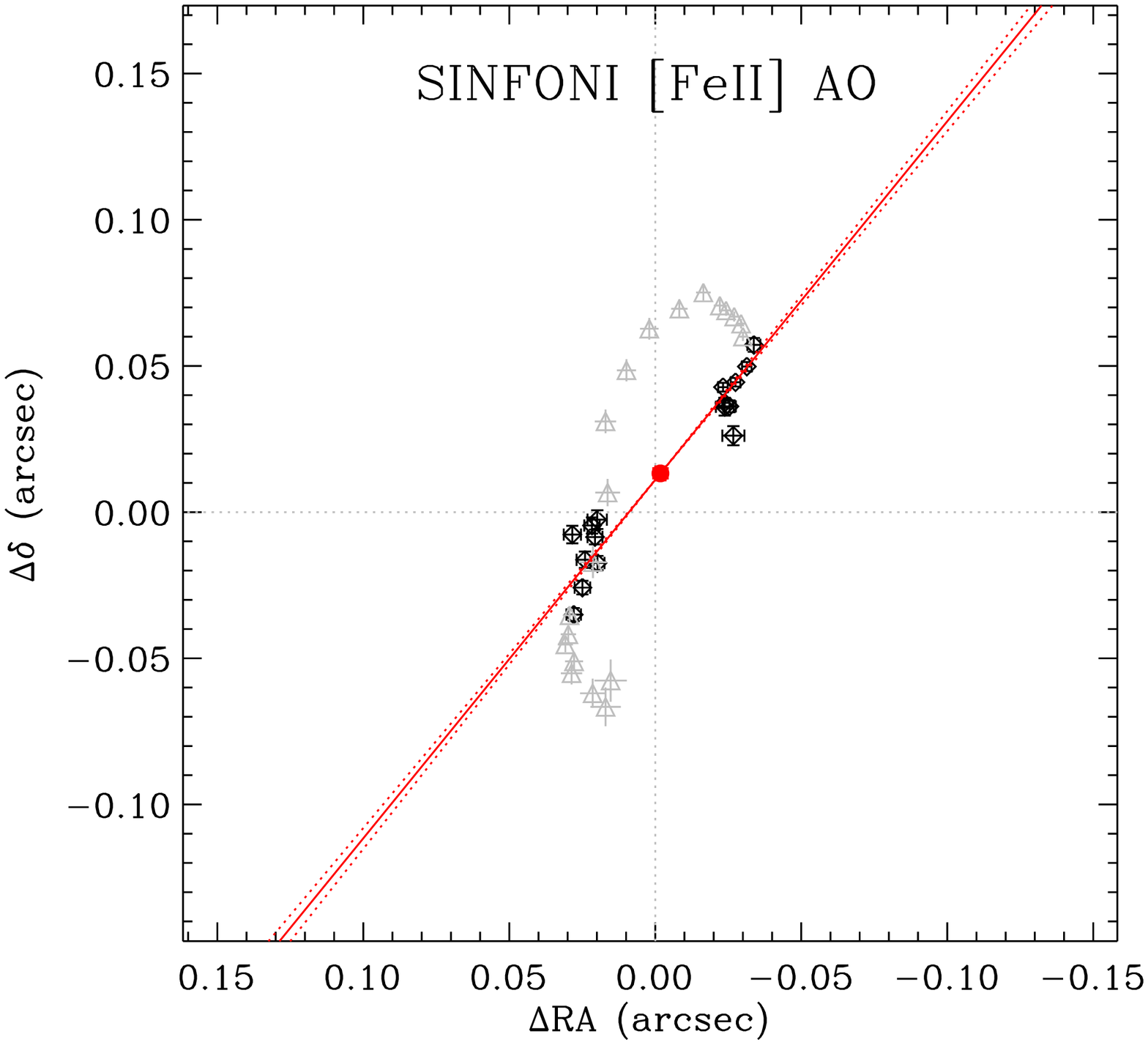}\\
  \caption{2D spectroastrometric map for the ISAAC longslit J band data (upper panels): Pa$\beta$ line (upper left panel) and [Fe\,II] line (upper right panel). Spectroastrometric map for the SINFONI H band data (lower panels): [Fe\,II] line for the seeing limited observation (lower left panel) and AO assisted [Fe\,II] line (lower right panel). The red point marks the inferred BH position. The red solid line represents the line of nodes of the disk obtained with a linear fit of the HV points, as described in the text. The dotted red lines represent the $1\sigma$ uncertainties on the line of nodes position angle. All boxes have the same angular dimension on the plane of sky ($\sim0.3\arcsec\times0.3\arcsec$).}  \label{fig10}\label{fig11}
  \end{figure*}
  
\section{Integral field spectra: observations and data analysis}
\subsection{Data and spectra analysis}

We use available near infrared spectra of the nucleus of Centaurus A obtained with SINFONI at the ESO VLT (see \citealt{Neumayer:2007} for details). In particular we make use of H band spectra observed in seeing limited mode and H and K band spectra obtained with the assistance of the adaptive optics (AO) system MACAO.
The seeing of the observations as measured by the seeing monitor was $FWHM_V \sim0.5\arcsec$ (transformed to K band , $FWHM_K \sim0.38\arcsec$).  Seeing limited spectra use a pixel scale of $0.125\arcsec\times0.250\arcsec$ and cover a $8\arcsec\times8\arcsec$ field of view.
The SINFONI AO module used as guide star an R $\sim14$ mag star $36\arcsec$ southwest of the nucleus providing a spatial resolution of $\sim0.12\arcsec$ (FWHM).  AO assisted spectra use a pixel scale of $0.050\arcsec\times0.100\arcsec$ and cover a $3\arcsec\times3\arcsec$ field of view.
Spectral resolution is $R\sim4000$ for the K band and a slightly lower for the H band, $R\sim3000$. The total on-source exposure for the K band data cube was $13500 s$ whereas for H band was $3600 s$.
For all details of observations and data reduction the reader is referred to \citealt{Neumayer:2007} and \citealt{Cappellari:2009}.

All data cubes were continuum subtracted by fitting a power law function to the spectrum of each spatial pixel (with emission and absorption lines masked) which was then subtracted. As observed in sections 3 and 4.1 of Paper I, for this application of spectroastrometry it is mandatory to use continuum subtracted spectra. The focus is exclusively on gas kinematics and the presence of an underlying continuum can significantly alter the spectroastrometric measurement of the emission line gas by modifying the spatial light distribution in each velocity channel. In the following spectroastrometric analysis we will use [FeII] observed in the H band ($1.6468\mu m$) and the $\mathrm{H_2}$ line observed in K band ($2.1259\mu m$). Wavelengths are then converted in velocity using as reference (zero velocity) the rest frame [Fe\,II] and $\mathrm{H_2}$ wavelengths (respectively $1.6435\mu m$ and $2.1213\mu m$).

\subsection{The spectroastrometric map of the source}

As observed in section 6 of Paper I, the extension of the spectroastrometric technique to integral field spectra is straightforward, and deriving a 2d spectroastrometric map becomes trivial due to the 2d spatial coverage of integral field spectra. The analysis of data now reduces to fitting a 2d Gaussian to each channel map in turn, yielding the X, Y positions of the photocenters as a function of velocity, i.e.~the spectroastrometric map. Therefore, we can directly derive the 2d spectroastrometric map from the continuum subtracted SINFONI data cubes, overcoming the problems related to the uncertainties in slit positioning with respect to the galaxy nucleus.

To select the HV range in the case of integral field data we also make use of the widths of the 2d Gaussians fitted to each velocity channel map. As explained in sect.~3 and appendix B of Paper I we search for unresolved spatial emission. In Fig. \ref{fig09} we show as an example the FWHMs of the fitted 2d Gaussian\footnote{Actually we fit a non circularly symmetric 2D Gaussian function. In Fig.~\ref{fig09} we show the minimum of the two FWHM values along the proper axis.} for the  seeing limited [FeII] data in the H band where the estimated seeing is $\sim0.4\arcsec$. We can observe an evident peak in the FWHM in the LV range due to the presence of spatially resolved  emission.  As previously, we identify the HV range by considering FWHMs lower than $1.1$ times the spatial resolution ($\sim0.4\arcsec$). The resulting range in this particular case is $v\lesssim250$km$/s$ and $v\gtrsim950$km$/s$.

 In the lower panels of Fig.~\ref{fig10} we show the derived spectroastrometric maps for the H band [Fe\,II] data (both AO assisted and not) and for the K band $\mathrm{H_2}$ AO assisted data. 
Typical uncertainties in the light centroid positions are of the order of $\sim0.01\arcsec$ for the seeing limited data (spatial resolution $\sim0.4\arcsec$ and pixel scale $0.125\arcsec\times0.250\arcsec$), and a factor $\sim2.5$ lower for the AO assisted data ($\sim0.004\arcsec$ with spatial resolution of $\sim0.12\arcsec$ and pixel scale of $0.05\arcsec\times0.10\arcsec$). It should also be noticed that all LV points lie outside the line of nodes, as expected.

 As previously explained, these 2d spectroastrometric maps can now be used to obtain the direction of the disk line of nodes and a first estimate of the BH position on the plane of the sky (see Table \ref{tab1}). The derived BH positions are consistent within $\sim0.02\arcsec$. The origin of the sky plane corresponds to the continuum peak position measured directly on the SINFONI datacubes with uncertainties of $\sim0.006\arcsec$ for the seeing limited data and $\sim0.002\arcsec$ for the AO data. Taking into account these systematic errors in setting the origin of the map and the typical uncertainties of the BH position estimate, BH positions are consistent among each other.


We also note that the derived $\theta_{LON}$ values for the [Fe\,II] line observed with or without the AO are consistent within $\sim8^{\circ}$. The differences between AO and seeing limited observations might also originate because of a warping in the gas disk which results in different orientations of the line of nodes at different spatial scales (see also \citealt{Neumayer:2007}).

From Fig.~\ref{fig10} we can also observe that the BH position derived from ISAAC J band spectra are shifted by $\sim0.08\arcsec$ with respect to that derived from SINFONI H band spectra. This effect can be accounted from the fact that the continuum peak positions measured in J and H band (i.e.~the maps coordinates origin) can be different due to the different effect of dust reddening on the two bands.

\subsection{Estimate of the BH mass from the spectroastrometric map} 

Here we estimate the BH mass using the spectroastrometric map as described in Sect.~5 of Paper I.
The application is exactly the same as in the case of the longslit spectra presented in section \ref{s44}, because in both cases we use as input the 2d spectroastrometric map (i.e. the positions of the light centroids in the plane of sky $x_{ch}, y_{ch}$ as a function of the corresponding channel velocity $V_{ch}$).
  \begin{figure*}[ht!]
  \centering
  \includegraphics[width=0.49\linewidth]{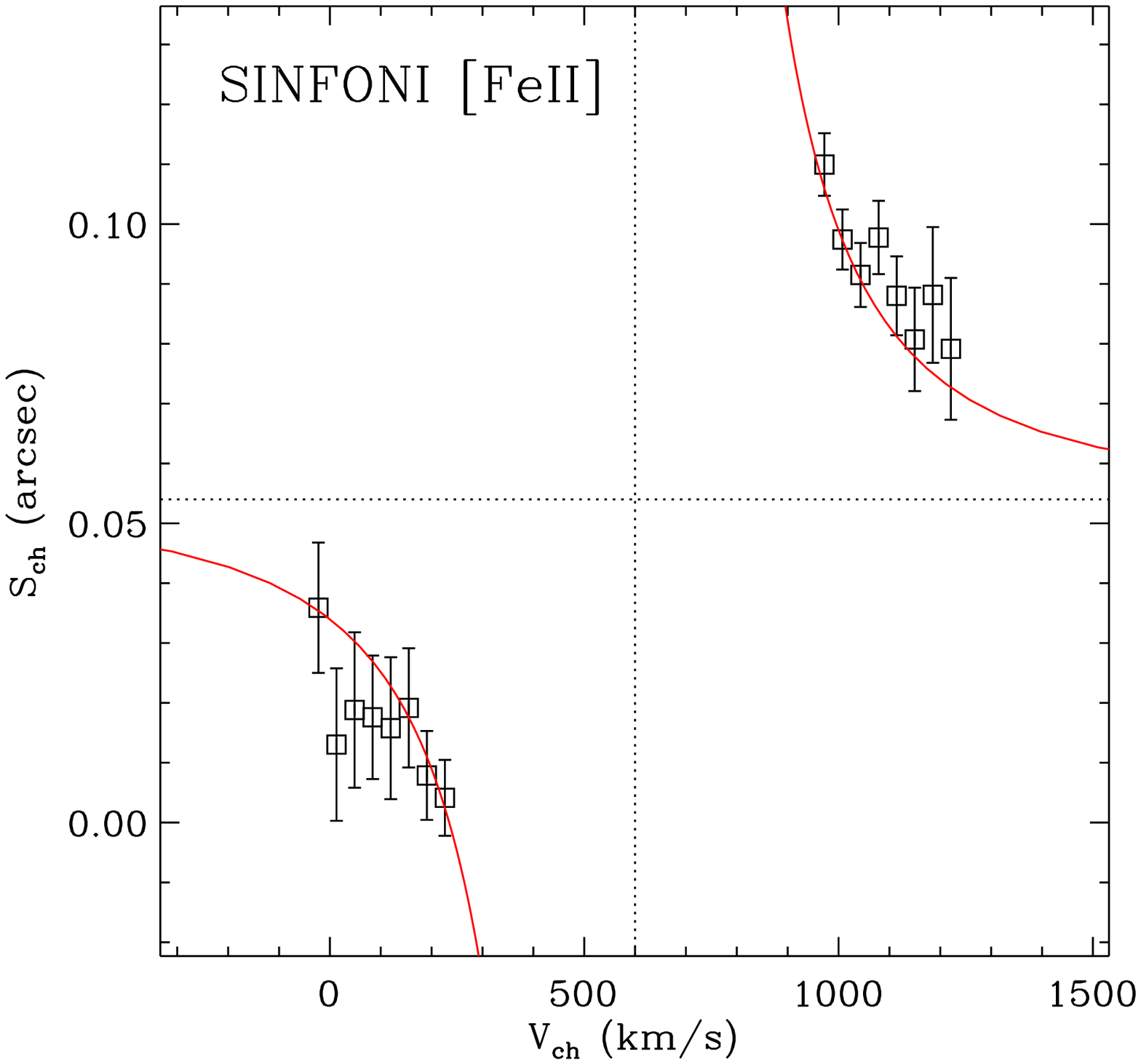}
  \includegraphics[width=0.49\linewidth]{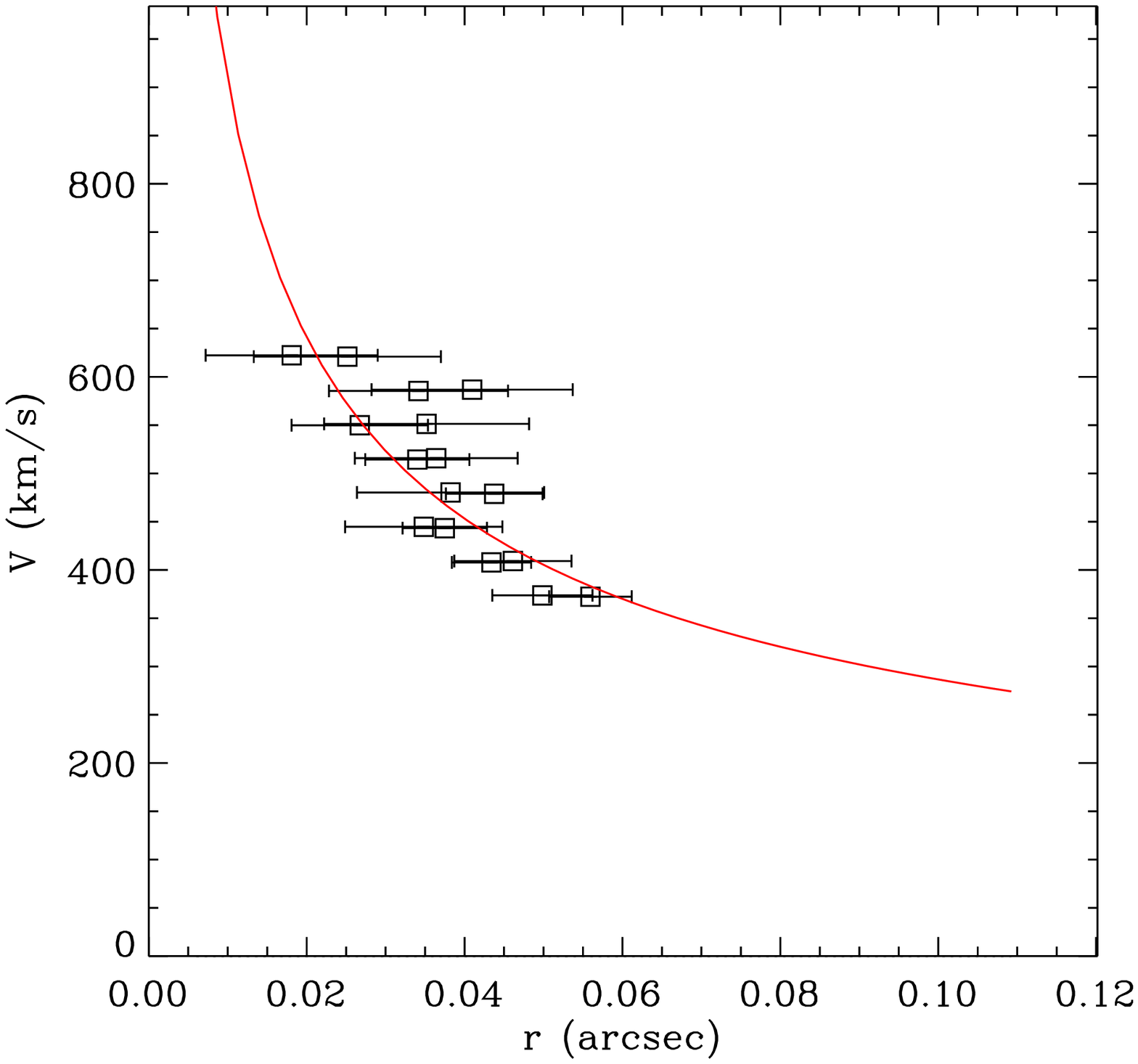}
  \includegraphics[width=0.49\linewidth]{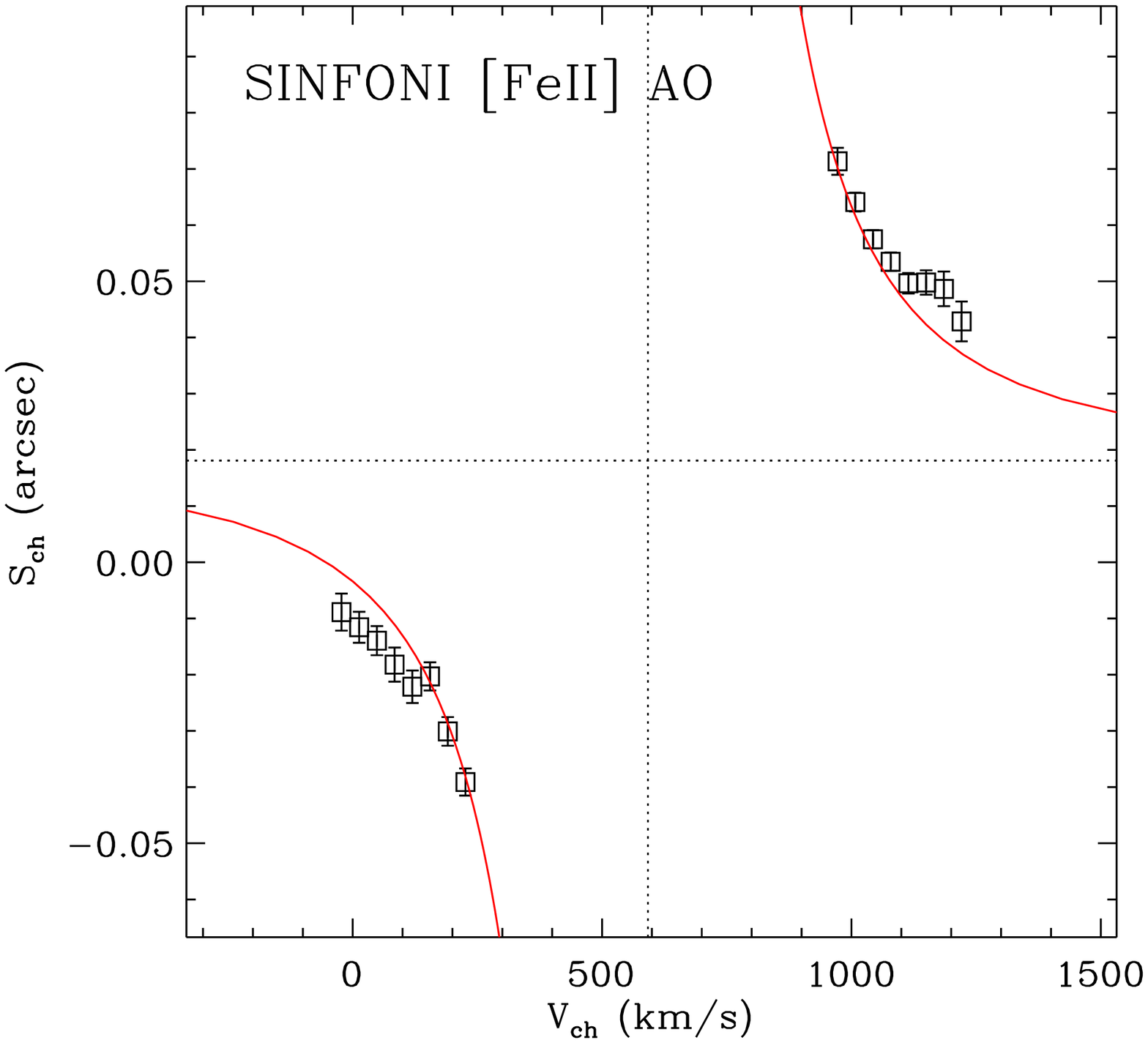}
  \includegraphics[width=0.49\linewidth]{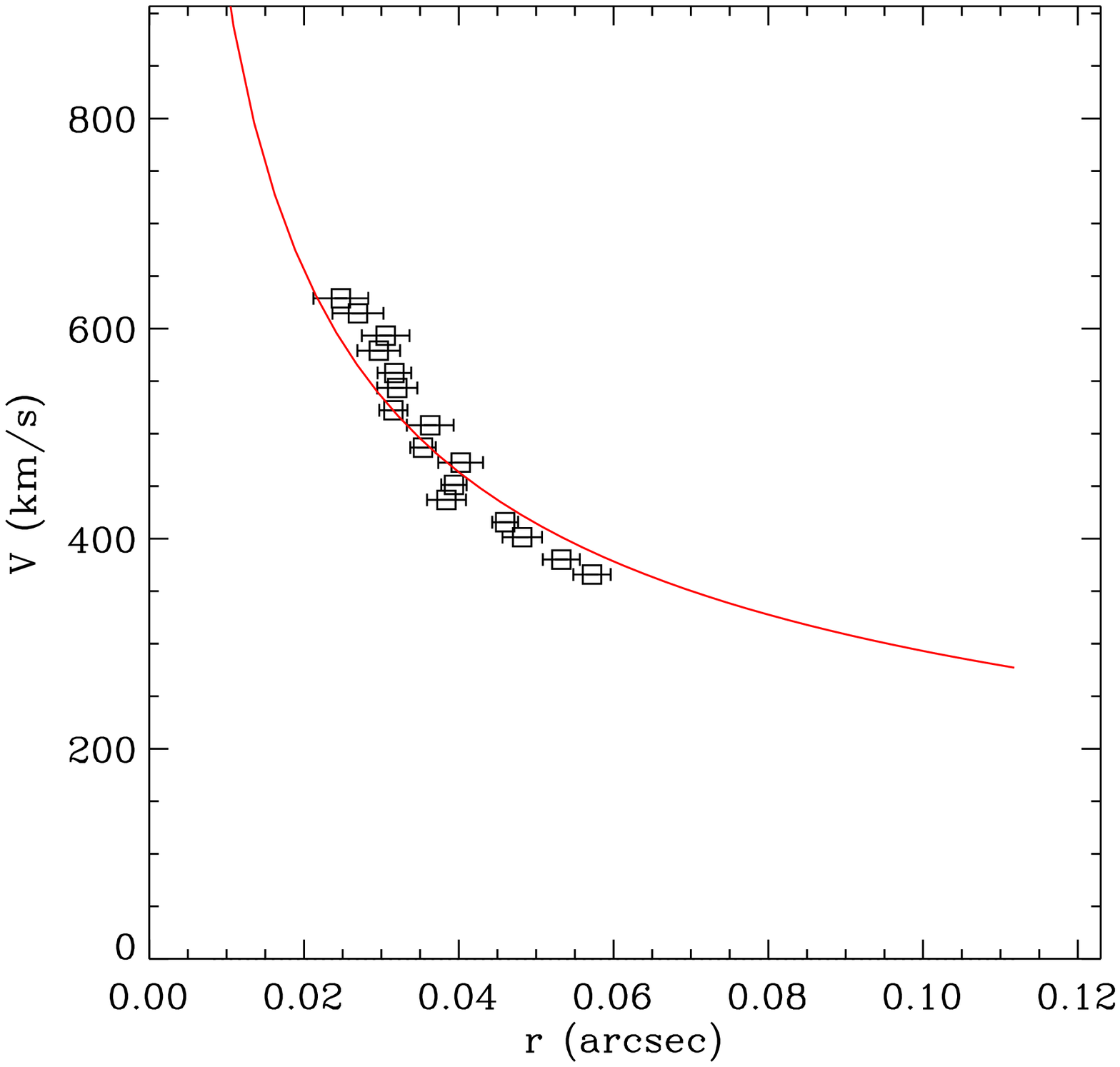}
     \caption{Results of the fit for the SINFONI seeing limited [Fe\,II] data (upper panels) and SINFONI AO assisted [Fe\,II] data (lower panels). The left panels show the line of nodes projected rotation curve $S_{ch}$ vs.  $V_{ch}$. The horizontal and vertical dotted lines denote respectively the BH line of nodes coordinate $S_0$ and the systemic velocity $V_{sys}$. The right panels show the $r=|S_{ch}-S_0|$ vs. $V=|V_{ch}-V_{sys}|$ rotation curve. The solid red lines represent the curves expected from the model.}
        \label{fig12}
  \end{figure*}

We have performed the fit of the data from the three spectroastrometric maps of Fig.~\ref{fig10}. Fit results are tabulated in Table \ref{tab1} and presented graphically in Fig.~\ref{fig12} where we plot either the $r=|S_{ch}-S_0|$ vs. $V=|V_{ch}-V_{sys}|$ rotation curve and the line of nodes projected rotation curve (e.g.  $S_{ch}$ vs. $V_{ch}$). The solid red lines represent the model rotation curves ($r$ vs. $|\bar{V}_{ch}-V_{sys}|$ and $S_{ch}$ vs. $\bar{V}_{ch}$).

\begin{table*}
   \caption[!h]{Fit Results.}
   \label{tab1}
   \centering
   \begin{tabular}{l c c c c c}
     \hline
     \noalign{\smallskip}
     Parameter & \multicolumn{5}{c}{Best fit value$\pm$error} \\
     \noalign{\smallskip}
     \hline
     \noalign{\smallskip}
       & \multicolumn{3}{c}{ISAAC data} & \multicolumn{2}{c}{SINFONI data} \\
     \noalign{\smallskip}
     & [Fe\,II] line fit & $Pa_\beta$ line fit & both lines fit & NO AO [Fe\,II] line fit & AO [Fe\,II] line fit \\
     \noalign{\smallskip}
     \hline
     \noalign{\smallskip}
     $\theta_{LON}\ \ \ [^{\circ}]\ ^{(1)}$      & $-20.3\pm1.2$  & $-18.3 \pm1.8$ & $-19.8 \pm1.0$ & $-47\pm2$  & $-39.2 \pm0.8$ \\
     \noalign{\smallskip}
     \hline
     \noalign{\smallskip}
     $log_{10}(M_{BH}\,sin^2i/M_{\astrosun})$ & $7.39\pm 0.02$ & $7.39\pm 0.03$ & $7.39\pm 0.02$ & $7.51\pm 0.06$ & $7.53\pm 0.01$ \\
     $log_{10}(M/L\,sin^2i/M_{\astrosun})$ & $-10.1\pm 0.0\ ^{(2)}$ & $-9.6\pm0.0\ ^{(2)}$ & $-7.9\pm0.0\ ^{(2)}$ & $-14.5\pm 0.0\ ^{(2)}$ & $-9.1\pm0.0\ ^{(2)}$ \\
     $S_0\ \ \ [\arcsec]$                    & $0.105\pm0.008$  & $0.103\pm0.009$ & $0.104\pm0.006$ & $0.054\pm0.006$  & $0.018\pm0.003$ \\
     $V_{sys}\ \ \ [km/s]$                              & $593.5\pm15.6$ & $584.8 \pm17.4$ & $591.4 \pm9.7$ & $599\pm33$ & $592\pm19$ \\
     $\Delta_{sys}\ \ \ [km/s]$                             & $11$ & $0$ & $0$  & $0$ & $19$ \\
     \noalign{\smallskip}
     \noalign{\smallskip}
     $\chi^2_{red} \ \ \ (\chi^2/D.O.F.)$ & $1.01\ \ (9.11/9)$ & $0.82\ \ (8.20/10)$ & $0.93\ \ (21.49/23)$& $0.55\ \ (6.59/12)$ & $1.02\ \ (12.2/12)$ \\
     \noalign{\smallskip}
     \hline
     \noalign{\smallskip}
     $x_{BH}\ \ \ [\arcsec]$        & $0.012\pm 0.006$ & $-0.003 \pm 0.007$ & $0.007 \pm 0.005$ & $-0.001\pm 0.007$ & $-0.002 \pm 0.002$ \\
     $y_{BH}\ \ \ [\arcsec]$        & $0.099 \pm 0.008$ & $0.098 \pm 0.008$ & $0.098 \pm 0.005$& $0.037 \pm 0.005$ & $0.014 \pm 0.002$ \\
     \noalign{\smallskip}
     \hline
   \end{tabular} 
\tablefoot{\tablefoottext{1}{Best fit parameter estimated in the fit of the line of nodes.}\tablefoottext{2}{Parameter not constrained from the fit.}}
\end{table*}

From the [FeII] data we estimate a value of the BH mass of $\log_{10}(M_{BH}sin^2i/M_{\astrosun}) = 7.5$, perfectly consistent between AO-assisted and seeing limited (tab.~\ref{tab1}) whereas from the $\mathrm{H_2}$ line we obtain a value ($\sim0.2$~dex) larger. Note that the higher accuracy of the light centroid positions of the spectroastrometric map for the AO assisted [FeII] data with respect to the seeing limited ones results in a lower uncertainty on the $M_{BH}sin^2i$ best fit value. 

$M/L$ is not constrained from the fit even with SINFONI data. As observed in section \ref{s44}, the radius of the BH sphere of influence for Centaurus A is $r_{BH}\simeq14.9pc$ corresponding to an apparent dimension of $\sim0.9\arcsec$; here we are studying the rotation curve at $\sim1/20$ smaller scales where the contribution  of the stellar mass  to the gravitational potential is negligible.

An impressive result of the spectroastrometric method is that the minimum radii at which we can probe the rotation curve are $\sim 25 mas$ for seeing limited data and $\sim20 mas$ for AO assisted data. The latter are only slightly smaller but have a much better positional accuracy, as shown in Fig.~\ref{fig12}. 
These values are $\sim1/16$ and $\sim1/6$, respectively, of the spatial resolution ($\sim0.4\arcsec$ for the seeing limited and $\sim0.12\arcsec$ for the AO assisted observations) and correspond to distances from the BH of $\sim0.42pc$ and $\sim 0.35pc$, respectively, $\sim1/40$ of the radius of the BH sphere of influence. This is a clear demonstration of the great potentials of spectroastrometry in overcoming the spatial resolution limit. 
\begin{table*}
   \caption[!h]{Mass estimates for Centaurus A}
   \label{tab2}
   \centering
   \begin{tabular}{l c}
     \hline
     \noalign{\smallskip}
        ``Classical'' method applications & $log_{10}(M_{BH}\,sin^2i/M_{\astrosun})$ \\
     \noalign{\smallskip}
     \hline
     \noalign{\smallskip}
     ISAAC J band [Fe\,II] and $Pa_\beta$ lines \citep{Marconi:2006} ($i=25^{\circ}$) & $7.39\pm 0.04$\\
     NACO H band AO [Fe\,II] line \citep{Haring-Neumayer:2006} ($i=45^{\circ}$) & $7.48^{+0.04}_{-0.06}$\\
     CIRPASS  J band $Pa_\beta$ line \citep{Krajnovic:2007} ($i=25^{\circ}$) & $7.2^{+0.1}_{-0.3}$\\
     SINFONI K band AO  $\mathrm{H_2}$ line \citep{Neumayer:2007} ($i=34^{\circ}$) & $7.1\pm0.1$\\
     \noalign{\smallskip}
     \hline
     \noalign{\smallskip}
      Spectroastrometric method applications & \\
     \noalign{\smallskip}
     \hline
     \noalign{\smallskip}
     ISAAC J band [Fe\,II] line & $7.39\pm 0.02$\\
     ISAAC J band $Pa_\beta$ line & $7.39\pm 0.03$\\
     ISAAC J band [Fe\,II] and $Pa_\beta$ simultaneous fit & $7.39\pm 0.02$\\
     SINFONI H band no AO [Fe\,II] line & $7.51\pm0.06$\\
     SINFONI H band AO  [Fe\,II] line & $7.52\pm 0.01$\\
     \noalign{\smallskip}
     \hline
   \end{tabular} 
\end{table*}

\subsection{SINFONI $\mathrm{H_2}$ line spectra}
As described above, we also have available AO assisted SINFONI spectra in K band where the molecular hydrogen line ($\mathrm{H_2}$) is observed. \cite{Neumayer:2007} used this emission line to study the nuclear gas kinematics and to estimate the BH mass. They find that this line spectrum has a good S/N and spatial resolution and that the derived gas kinematics are well explained by a rotating disk. In contrast the kinematics from the [Fe\,II] H band spectra clearly show disturbances from the presence of the jet.

However, $\mathrm{H_2}$ proved to be of lower quality for the purpose of spectroastrometric analysis. In particular, the line emission is spatially well extended and the S/N in the HV line wings is much lower with respect to the [Fe\,II] line. At first sight this might seem surprising given the high quality of the data obtained by \citealt{Neumayer:2007}, but this is due to the different features of the spectroastrometric method. In the spectroastrometric application, as extensively discussed in Paper I, we search for a spatially unresolved emission in the HV wings of the line as a ``signature'' of the gas kinematics of the BH gravitational potential and discard all the spatially resolved emission in the LV range. On the contrary, in standard gas dynamical studies, it is important to concentrate on the spatially resolved LV range of the line spectrum, and particular gas kinematics features like the presence of inflows, outflows or jets can be detected in this conditions.

\begin{figure}[ht!]
  \centering
  \includegraphics[width=0.95\linewidth, trim= 20 0 2 0]{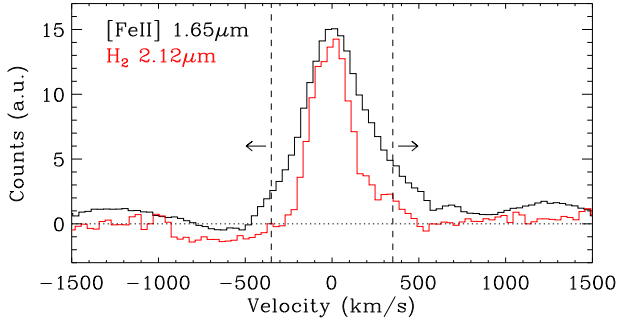}
     \caption{Comparison between the [Fe\,II] and $\mathrm{H_2}$ line spectra extracted from a circular aperture of $\sim0.75\arcsec$ centered on the continuum peak (from SINFONI AO assisted observations). Overplotted the HV range limits used for the [Fe\,II] analysis. The $0\,km/s$ velocity correspond to the respective central line wavelengths and the counts are rescaled to approximately match the two line peaks.}
        \label{fig13}
  \end{figure}

In Fig.~\ref{fig13} we show the comparison between the line profiles of the [Fe\,II] and $\mathrm{H_2}$ lines. Both spectra extracted from the respective SINFONI datacubes (AO assisted observations) from a circular aperture of $\sim0.75\arcsec$ centered on the continuum peak. Clearly the $\mathrm{H_2}$ line spectrum has almost no signal in the HV range compared to the [Fe\,II] line. For this reason the application of our method results in a poor quality spectroastrometric map. On the other hand in the present work we find the SINFONI [Fe\,II] spectra very useful and it seems our results are not dominated by the contamination of the jet as observed for the [FeII] kinematics by \citealt{Neumayer:2007}. This is because we are concentrating on the unresolved HV gas emission that comes from the inner region of the nuclear disk and that seems not to be strongly influenced by the presence of the jet.

Summarizing, [FeII] is detected both on large spatial scales (those probed with the classical kinematical analysis) and on small spatial scales (those probed with the spectroastrometric analysis). On large spatial scales and low velocities, the [FeII] kinematics is likely affected by the presence of the jet, while this is not the case for $\mathrm{H_2}$ (see \citealt{Neumayer:2007} for more details).
On small spatial scales and high velocities, the [FeII] kinematics is little affected by the presence of the jet, if any, while $\mathrm{H_2}$ emission is absent and therefore cannot be used for spectroastrometry. This different behavior of [FeII] emitting gas is might be related to the warping of the disk, which at small spatial scales tends to be perpendicular to the jet axis.

\section{Discussion and conclusions}
We have obtained new mass measurements for the nuclear BH of the Centaurus A galaxy by applying the spectroastrometric method to longslit and both seeing limited and AO corrected IFU spectra. Compared to the standard method based on rotation curves analysis, the spectroastrometric method is much simpler from the modeling point of view; it only requires the determination of the ``spectroastrometric map" and is relatively insensitive by the problems that plague the standard approach based on rotation curves like, e.g., the effect of beam smearing, the intrinsic flux distribution of the line, and the biases due to the slit positioning in longslit observations.

With our proposed spectroastrometric approach we can derive two-dimensional plane-of-the-sky spectroastrometric maps  of the source characterized by accuracies of position measurements much lower than the spatial resolution of the observations. The mean accuracies of our light centroid position estimates are $\sim0.01\arcsec$ for the ISAAC data and seeing limited SINFONI data ($\sim1/40$ of the spatial resolution) and $\sim0.004\arcsec$ for AO assisted SINFONI data ($\sim1/30$ of the spatial resolution).
The position angles of the disk line of nodes  estimated from those maps are $\sim-19^{\circ}$ for the two ISAAC maps (consistent within $\sim2^{\circ}$). The estimates from SINFONI  maps are $\sim20^{\circ}$ offset from the previous. However  the  SINFONI [FeII] observations  (seeing limited and AO assisted) provides line of nodes PA estimates consistent within $\sim8^{\circ}$.

The differences between the various estimates of the disk line of nodes PA might be due to the different ways in which the spectroastrometric maps are constructed, i.e.~combining three long-slit spectroastrometric curves for ISAAC data and directly from the datacube for SINFONI data. Moreover there are many clear indications that the Centaurus A nuclear disk is warped (see \citealt{Neumayer:2007}) and therefore the different spectroastrometric maps might probe the average disk line of nodes at different spatial scales. In any case, the derived rotation curves and hence the BH mass estimates are consistent and not affected by the differences in PA.

One possible cause of concern could be the support provided against gravity by turbulent motions. However \cite{Marconi:2006} showed that the line widths of ionized lines and Pa$\beta$ are consistent with unresolved rotation and therefore there is no indication for turbulent pressure. Moreover the BH mass determination based on $\mathrm{H_2}$ by \cite{Neumayer:2007} which includes turbulent support is similarly consistent with the result by \cite{Marconi:2006} (once the different disk inclinations are taken into account, see Fig.~\ref{fig01}) indicating that its effect is small.

The most important result in this work is the demonstration of the capability of spectroastrometry to overcome the spatial resolution limit and estimate BH masses in a simple and neat way.
The minimum distance from the BH at which we can probe the gas rotation curve is $\sim50mas$ for the ISAAC data ($\sim1/10$ of the spatial resolution) and $\sim20mas$ for the SINFONI data ($\sim1/15$ of the spatial resolution for the seeing limited data and $\sim1/6$ for the AO assisted data). In the case of Centaurus A, these corresponds to $\sim1/30$ and $\sim1/50$, respectively, of the radius of the BH sphere of influence indicating that is is possible to probe deep in the BH potential well, where the contribution from the mass in stars is negligible.

In table \ref{tab2} we compare $M_{BH}sin^2i$ values resulting from recent applications of the classical rotation curve method to various data sets of Centaurus A (including those analyzed here) with our results from the application of the spectroastrometric method. From Table \ref{tab2} several conclusion can be reached.
\begin{itemize}
\item
Our new simple method based on spectroastrometry is in excellent agreement with the classical method based on the rotation curves, at least when comparing the results obtained from the same dataset (cf.~the ISAAC \citealt{Marconi:2006} data). This is a fundamental indication for the robustness of our new method: using the same dataset we can apply indifferently the classical and the spectroastrometrical method obtaining perfectly consistent results.
\item 
When applying the two methods to different datasets but with the same target line we obtain consistent estimates (cf. the $M_{BH}$ estimate by \citealt{Haring-Neumayer:2006} targeting the H band [FeII] line  and the one presented in this paper, and that by \citealt{Krajnovic:2007} based on the J band $Pa_\beta$, with our own which is within $\sim0.2$~dex).
\item 
In general all the measurements reported in Table \ref{tab2} are consistent within $\pm0.2$~dex and these differences are mainly due to the different data sets and target lines and not from the application of a particular method.
\item 
The application of the spectroastrometric method to different data type (IFU and longslit) give consistent result (within only $\sim0.1$~dex) and this demonstrates the versatility of our method. The application to IFU data is obviously much simpler but this agreement also show that our method of ``reconstruction" of the 2d map from multiple longslit spectroastrometry is correct.
\item
The application of the method to IFU data with and without AO also produces consistent results at similar spatial scales. This clearly demonstrates how spectroastrometry is much less sensitive to spatial resolution than the classical method. As expected, the accuracy of measurement positions without AO is worse. This is due to the fact that for each velocity channel we recover the centroid position by fitting a two dimensional Gaussian function and the width of this function clearly decreases with increasing spatial resolution of the data. Since this function is centro-symmetric we can always recover the correct center position. Increasing the width of the function only makes the center position uncertainty larger but does not alter its position.
\end{itemize}

A typical feature of spectroastrometry is its insensitivity to disk inclination. In our fit, in fact, mass and disk inclination are coupled. This is because coordinates along the line of nodes ($S_{ch}$) do not depend on $i$, which only appears as an unavoidable projecting factor on the velocity. Therefore our method can return only a $M_{BH}\,sin^2i$ value with the need of assuming an inclination value to derive the value of the mass. 
In this paper we decided to present only the $M_{BH}\,sin^2i$ values and to move the discussion on the $M_{BH}$ values to a discussion on the $i$ values estimated or assumed by the various authors. For completeness we report in Table \ref{tab2} the inclinations estimated or assumed in those previous works.
Summarizing, we have applied the spectroastrometric method presented in Paper I to several datasets of the nucleus of the Centaurus A galaxy obtaining $M_{BH}$ estimates which are consistent with the classical method based on rotation curves.

In conclusion, the application to Centaurus A has shown that the spectroastrometric method is much simpler and straightforward than the classical rotation curves method and provides the major advantage of enabling spatial scale much smaller than the spatial resolution to be probed. It is thus possible to obtain more robust and accurate BH mass measurements where the classical method fails. 
Therefore the spectroastrometric method can be used to extend the measured $M_{BH}$ range to smaller BHs in the local universe and to similar BHs in more distant galaxies than is currently possible.
Moreover this method can be applied to any type of long-slit or integral field spectra without any particular regard for instrument or wavelength range. A very interesting and promising application, for example, could be in the sub-mm high spatial resolution data provided by ALMA.

So far the spectroastrometric method has been applied independently from the classical method. However as discussed in Paper I, it is clear that the spectroastrometric and ``classical'' rotation curves are complementary and orthogonal descriptions of the position velocity diagram. Therefore, a future development of this method will be its application in combination to the classical method based on rotation curves. This will also allow us to constrain the disk inclination and thus remove mass-inclination degeneracy.

\begin{acknowledgements}
We would like to thank the anonymous referee for a useful and constructive report on this paper. We also acknowledge financial support from the Italian National Institute for Astrophysics by INAF CRAM 1.06.09.10
\end{acknowledgements}

\bibliographystyle{aa} 

\end{document}